\documentclass[%
 reprint,
superscriptaddress,
 amsmath,amssymb,
 aps,
]{revtex4-1}

\usepackage{graphicx}
\usepackage{dcolumn}
\usepackage{bm}

\begin{document}

\preprint{APS/123-QED}

\title{Measurement of the inclusive $\nu_{\mu}$ charged current cross section\\on iron and hydrocarbon in the T2K on-axis neutrino beam}


\newcommand{\INSTC}{\affiliation{University of Alberta, Centre for Particle Physics, Department of Physics, Edmonton, Alberta, Canada}}
\newcommand{\INSTEE}{\affiliation{University of Bern, Albert Einstein Center for Fundamental Physics, Laboratory for High Energy Physics (LHEP), Bern, Switzerland}}
\newcommand{\INSTFE}{\affiliation{Boston University, Department of Physics, Boston, Massachusetts, U.S.A.}}
\newcommand{\INSTD}{\affiliation{University of British Columbia, Department of Physics and Astronomy, Vancouver, British Columbia, Canada}}
\newcommand{\INSTGA}{\affiliation{University of California, Irvine, Department of Physics and Astronomy, Irvine, California, U.S.A.}}
\newcommand{\INSTI}{\affiliation{IRFU, CEA Saclay, Gif-sur-Yvette, France}}
\newcommand{\INSTGB}{\affiliation{University of Colorado at Boulder, Department of Physics, Boulder, Colorado, U.S.A.}}
\newcommand{\INSTFG}{\affiliation{Colorado State University, Department of Physics, Fort Collins, Colorado, U.S.A.}}
\newcommand{\INSTFH}{\affiliation{Duke University, Department of Physics, Durham, North Carolina, U.S.A.}}
\newcommand{\INSTBA}{\affiliation{Ecole Polytechnique, IN2P3-CNRS, Laboratoire Leprince-Ringuet, Palaiseau, France }}
\newcommand{\INSTEF}{\affiliation{ETH Zurich, Institute for Particle Physics, Zurich, Switzerland}}
\newcommand{\INSTEG}{\affiliation{University of Geneva, Section de Physique, DPNC, Geneva, Switzerland}}
\newcommand{\INSTDG}{\affiliation{H. Niewodniczanski Institute of Nuclear Physics PAN, Cracow, Poland}}
\newcommand{\INSTCB}{\affiliation{High Energy Accelerator Research Organization (KEK), Tsukuba, Ibaraki, Japan}}
\newcommand{\INSTED}{\affiliation{Institut de Fisica d'Altes Energies (IFAE), Bellaterra (Barcelona), Spain}}
\newcommand{\INSTEC}{\affiliation{IFIC (CSIC \& University of Valencia), Valencia, Spain}}
\newcommand{\INSTEI}{\affiliation{Imperial College London, Department of Physics, London, United Kingdom}}
\newcommand{\INSTGF}{\affiliation{INFN Sezione di Bari and Universit\`a e Politecnico di Bari, Dipartimento Interuniversitario di Fisica, Bari, Italy}}
\newcommand{\INSTBE}{\affiliation{INFN Sezione di Napoli and Universit\`a di Napoli, Dipartimento di Fisica, Napoli, Italy}}
\newcommand{\INSTBF}{\affiliation{INFN Sezione di Padova and Universit\`a di Padova, Dipartimento di Fisica, Padova, Italy}}
\newcommand{\INSTBD}{\affiliation{INFN Sezione di Roma and Universit\`a di Roma ``La Sapienza'', Roma, Italy}}
\newcommand{\INSTEB}{\affiliation{Institute for Nuclear Research of the Russian Academy of Sciences, Moscow, Russia}}
\newcommand{\INSTHA}{\affiliation{Kavli Institute for the Physics and Mathematics of the Universe (WPI), Todai Institutes for Advanced Study, University of Tokyo, Kashiwa, Chiba, Japan}}
\newcommand{\INSTCC}{\affiliation{Kobe University, Kobe, Japan}}
\newcommand{\INSTCD}{\affiliation{Kyoto University, Department of Physics, Kyoto, Japan}}
\newcommand{\INSTEJ}{\affiliation{Lancaster University, Physics Department, Lancaster, United Kingdom}}
\newcommand{\INSTFC}{\affiliation{University of Liverpool, Department of Physics, Liverpool, United Kingdom}}
\newcommand{\INSTFI}{\affiliation{Louisiana State University, Department of Physics and Astronomy, Baton Rouge, Louisiana, U.S.A.}}
\newcommand{\INSTJ}{\affiliation{Universit\'e de Lyon, Universit\'e Claude Bernard Lyon 1, IPN Lyon (IN2P3), Villeurbanne, France}}
\newcommand{\INSTCE}{\affiliation{Miyagi University of Education, Department of Physics, Sendai, Japan}}
\newcommand{\INSTDF}{\affiliation{National Centre for Nuclear Research, Warsaw, Poland}}
\newcommand{\INSTFJ}{\affiliation{State University of New York at Stony Brook, Department of Physics and Astronomy, Stony Brook, New York, U.S.A.}}
\newcommand{\INSTGJ}{\affiliation{Okayama University, Department of Physics, Okayama, Japan}}
\newcommand{\INSTCF}{\affiliation{Osaka City University, Department of Physics, Osaka, Japan}}
\newcommand{\INSTGG}{\affiliation{Oxford University, Department of Physics, Oxford, United Kingdom}}
\newcommand{\INSTBB}{\affiliation{UPMC, Universit\'e Paris Diderot, CNRS/IN2P3, Laboratoire de Physique Nucl\'eaire et de Hautes Energies (LPNHE), Paris, France}}
\newcommand{\INSTGC}{\affiliation{University of Pittsburgh, Department of Physics and Astronomy, Pittsburgh, Pennsylvania, U.S.A.}}
\newcommand{\INSTFA}{\affiliation{Queen Mary University of London, School of Physics and Astronomy, London, United Kingdom}}
\newcommand{\INSTE}{\affiliation{University of Regina, Department of Physics, Regina, Saskatchewan, Canada}}
\newcommand{\INSTGD}{\affiliation{University of Rochester, Department of Physics and Astronomy, Rochester, New York, U.S.A.}}
\newcommand{\INSTBC}{\affiliation{RWTH Aachen University, III. Physikalisches Institut, Aachen, Germany}}
\newcommand{\INSTFB}{\affiliation{University of Sheffield, Department of Physics and Astronomy, Sheffield, United Kingdom}}
\newcommand{\INSTDI}{\affiliation{University of Silesia, Institute of Physics, Katowice, Poland}}
\newcommand{\INSTEH}{\affiliation{STFC, Rutherford Appleton Laboratory, Harwell Oxford,  and  Daresbury Laboratory, Warrington, United Kingdom}}
\newcommand{\INSTCH}{\affiliation{University of Tokyo, Department of Physics, Tokyo, Japan}}
\newcommand{\INSTBJ}{\affiliation{University of Tokyo, Institute for Cosmic Ray Research, Kamioka Observatory, Kamioka, Japan}}
\newcommand{\INSTCG}{\affiliation{University of Tokyo, Institute for Cosmic Ray Research, Research Center for Cosmic Neutrinos, Kashiwa, Japan}}
\newcommand{\INSTGI}{\affiliation{Tokyo Metropolitan University, Department of Physics, Tokyo, Japan}}
\newcommand{\INSTF}{\affiliation{University of Toronto, Department of Physics, Toronto, Ontario, Canada}}
\newcommand{\INSTB}{\affiliation{TRIUMF, Vancouver, British Columbia, Canada}}
\newcommand{\INSTG}{\affiliation{University of Victoria, Department of Physics and Astronomy, Victoria, British Columbia, Canada}}
\newcommand{\INSTDJ}{\affiliation{University of Warsaw, Faculty of Physics, Warsaw, Poland}}
\newcommand{\INSTDH}{\affiliation{Warsaw University of Technology, Institute of Radioelectronics, Warsaw, Poland}}
\newcommand{\INSTFD}{\affiliation{University of Warwick, Department of Physics, Coventry, United Kingdom}}
\newcommand{\INSTGE}{\affiliation{University of Washington, Department of Physics, Seattle, Washington, U.S.A.}}
\newcommand{\INSTGH}{\affiliation{University of Winnipeg, Department of Physics, Winnipeg, Manitoba, Canada}}
\newcommand{\INSTEA}{\affiliation{Wroclaw University, Faculty of Physics and Astronomy, Wroclaw, Poland}}
\newcommand{\INSTH}{\affiliation{York University, Department of Physics and Astronomy, Toronto, Ontario, Canada}}

\INSTC
\INSTEE
\INSTFE
\INSTD
\INSTGA
\INSTI
\INSTGB
\INSTFG
\INSTFH
\INSTBA
\INSTEF
\INSTEG
\INSTDG
\INSTCB
\INSTED
\INSTEC
\INSTEI
\INSTGF
\INSTBE
\INSTBF
\INSTBD
\INSTEB
\INSTHA
\INSTCC
\INSTCD
\INSTEJ
\INSTFC
\INSTFI
\INSTJ
\INSTCE
\INSTDF
\INSTFJ
\INSTGJ
\INSTCF
\INSTGG
\INSTBB
\INSTGC
\INSTFA
\INSTE
\INSTGD
\INSTBC
\INSTFB
\INSTDI
\INSTEH
\INSTCH
\INSTBJ
\INSTCG
\INSTGI
\INSTF
\INSTB
\INSTG
\INSTDJ
\INSTDH
\INSTFD
\INSTGE
\INSTGH
\INSTEA
\INSTH

\author{K.\,Abe}\INSTBJ
\author{J.\,Adam}\INSTFJ
\author{H.\,Aihara}\INSTCH\INSTHA
\author{T.\,Akiri}\INSTFH
\author{C.\,Andreopoulos}\INSTEH
\author{S.\,Aoki}\INSTCC
\author{A.\,Ariga}\INSTEE
\author{S.\,Assylbekov}\INSTFG
\author{D.\,Autiero}\INSTJ
\author{M.\,Barbi}\INSTE
% ...

\collaboration{The T2K Collaboration}\noaffiliation

\date{\today}

\begin{abstract}
We report a measurement of the $\nu_\mu$ inclusive charged current cross sections on iron and hydrocarbon in the T2K on-axis neutrino beam.
The measured inclusive charged current cross sections on iron and hydrocarbon averaged over the T2K on-axis flux with a mean neutrino energy of 1.51 GeV are
$(1.444\pm0.002(stat.)_{-0.157}^{+0.189}(syst.))\times 10^{-38}\mathrm{cm}^2/\mathrm{nucleon}$, and $(1.379\pm0.009(stat.)_{-0.147}^{+0.178}(syst.))\times 10^{-38}\mathrm{cm}^2/\mathrm{nucleon}$, respectively, and their cross section ratio is $1.047\pm0.007(stat.)\pm0.035(syst.)$.
These results agree well with the predictions of the neutrino interaction model, and thus we checked the correct treatment of the nuclear effect for iron and hydrocarbon targets in the model within the measurement precisions.

\end{abstract}

\pacs{Valid PACS appear here}
\maketitle


\section{Introduction}\label{sec:introduction}
The Tokai-to-Kamioka (T2K) experiment is a long baseline neutrino oscillation experiment \cite{t2k_nim}
whose primary goal is a study of the neutrino oscillations via the appearance of electron neutrinos and the disappearance of muon neutrinos.
An almost pure intense muon neutrino beam is produced at Japan Proton Accelerator Research Complex (J-PARC) in Tokai.
The proton beam impinges on a graphite target to produce charged pions, which are focused by three magnetic horns \cite{horn_ichikawa}. The pions decay
mainly into muon - muon-neutrino pairs during their passage through the 96-meter decay volume.
The neutrinos are measured by the near detectors (INGRID \cite{ingrid_nim} and ND280 \cite{p0d_nim,tpc_abgrall,fgd_amaudruz,ecal_jinst,smrd_nim}) in J-PARC and the far detector (Super-Kamiokande \cite{sk_detector}) in Kamioka, located 295km away from J-PARC.

A precise neutrino oscillation measurement requires good knowledge of neutrino interaction cross sections.
The neutrino charged current (CC) interaction is especially important for neutrino oscillation measurements
because the neutrino flavor is identifiable via the CC interaction.
Charged current neutrino-nucleon interactions at neutrino energies around 1 GeV have been studied in the past predominantly on deuterium targets \cite{barish, baker}.
Many modern neutrino oscillation experiments use heavier targets like carbon, oxygen and iron.
Nuclear effects are large for those targets and, consequently,
they cause large systematic uncertainties for the neutrino oscillation measurement in the case that there is no near detector or the
near and far detectors have different target material.
Therefore, it is important to measure and understand these interactions to minimize systematic uncertainties for the neutrino oscillation measurement.

In this paper, we present measurements of the inclusive muon neutrino charged current cross section on iron and hydrocarbon and their cross section ratio at neutrino energies around 1 GeV using the INGRID detector.
INGRID is located on the beam center axis and consists of 16 standard modules and an extra module called the Proton Module.
Iron (Fe) makes up 96.2\% of the target mass in the standard module, and hydrocarbon (CH) makes up 98.6\% of the target mass in the Proton Module.
Thus, the $\nu_{\mu}$ CC inclusive cross sections on Fe and CH are calculated from the number of selected CC events in one of the standard modules and the Proton Module respectively.
The $\nu_{\mu}$ CC inclusive cross section on Fe at neutrino energies above 3.5GeV was measured by the MINOS experiment \cite{minos_ccinc}, however the CC inclusive cross section around 1 GeV had never been measured.
Although the $\nu_{\mu}$ CC inclusive cross section on CH around 1 GeV was already measured by the T2K off-axis near detector ND280 \cite{t2k_ccinc} and other experiments\cite{nomad_ccinc, sciboone_ccinc}, the Proton Module can measure the cross section for higher energy neutrinos than the ND280 measurement, because the energy distribution of the on-axis neutrinos is higher than that of the off-axis neutrinos (the average energies of the on-axis and off-axis neutrinos are 1.51 GeV and 0.85 GeV, respectively).
We also measured the $\nu_{\mu}$ CC inclusive cross section ratio on Fe to CH using a central standard module and the Proton Module.
The central standard module and the Proton Module are on the central axis of the beam and are exposed to the same neutrino beam.
Thus, this cross section ratio can be measured very precisely, since many of the large systematic errors from uncertainties on the neutrino flux and neutrino interactions will be cancelled between the two detectors.
The CC inclusive cross section ratio on different target nuclei is expected to be differ from unity due to the difference in the ratio of neutrons and protons in the nuclei.
In addition, it will be affected by the nuclear effect especially in the low energy region.
Therefore, this measurement can provide a good test of the nuclear effect in the neutrino interaction model.
Recently, the MINER$\nu$A experiment measured the  cross section ratio at neutrino energies of 2--20 GeV \cite{minerva_ratio}. We can provide a result of the cross section ratio at a lower energy.

T2K collected data corresponding to $6.57\times10^{20}$ protons on target (POT) during the four run periods listed in Table~\ref{dataset}, with which $\nu_{\mu} \rightarrow \nu_e$ appearance was observed \cite{t2k_nue_app}.
During this time period, INGRID recorded more than 99.5\% of the delivered beam data.
A subset of data corresponding to $0.21\times 10^{20}$ POT from
Run 3 was collected with the magnetic horns operating
at 205 kA instead of the nominal value of 250 kA.
The Run 3 periods with the magnetic horns operating at 205 kA and 250 kA are referred to as Run 3b and Run 3c, respectively.
For the cross section measurement, data from Run 1, in which the Proton Module was not installed, and Run 3b are not used.
The total data set for the cross section measurement corresponds to $6.04\times10^{20}$ POT.

\begin{table}[htbp]
\begin{center}
  \caption{T2K data-taking periods and integrated
protons on target (POT). Data of Run 1 and Run 3b were not used for the cross section measurement.}
  \begin{tabular}{ccc}
    \hline\hline
     Run period & Dates & Integrated POT \\ 
    \hline
     (Run 1) & Jan. 2010 $-$ Jun. 2010 & $0.32\times10^{20}$\\ 
     Run 2 & Nov. 2010 $-$ Mar. 2011 & $1.11\times10^{20}$\\
     (Run 3b) & Mar. 2012 & $ 0.22\times10^{20}$\\
     Run 3c & Apr. 2012 $-$ Jun. 2012 & $1.37\times10^{20}$\\
     Run 4 & Oct. 2012 $-$ May. 2013 & $3.56\times10^{20}$ \\
    \hline\hline
  \end{tabular}
  \label{dataset}
  \end{center}
\end{table}

The remainder of this paper is organized as follows:
Details of the INGRID detector and Monte Carlo simulations are explained in Secs. \ref{sec:detector} and \ref{sec:mc}, respectively.
Section \ref{sec:event_selection} summarizes the CC-inclusive event selection.
The analysis method of the cross section measurement is described in Sec. \ref{sec:method}.
Section \ref{sec:error} describes the systematic errors.
The results and conclusions are given in Secs. \ref{sec:result} and \ref{sec:conclusion}, respectively.

\section{Detector configuration}\label{sec:detector}
The INGRID (Interactive Neutrino GRID) detector is an on-axis neutrino near detector located 280m downstream of the proton target.
It consists of 16 identical standard modules and an extra module called Proton Module.

\subsection{INGRID standard modules}
\begin{figure}[htbp]
  \begin{center}
  \includegraphics[width=50mm]{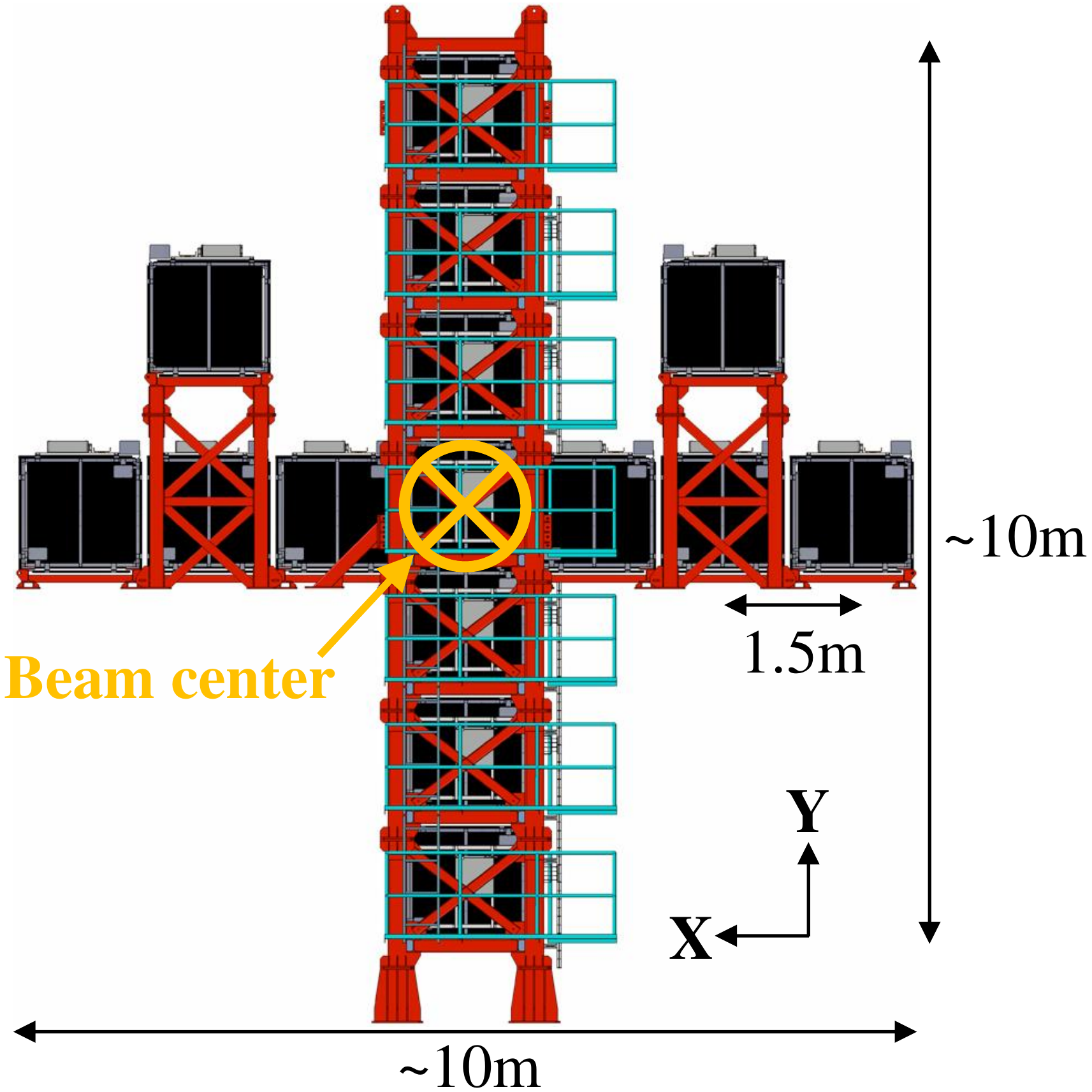}
    \caption{Overview of the 16 INGRID standard modules viewed from the beam upstream. The horizontal center module is hidden behind the vertical center module.}
  \label{ingrid_overview}
  \end{center}
\end{figure}

\begin{figure}[htbp]
  \begin{center}
  \includegraphics[width=84mm]{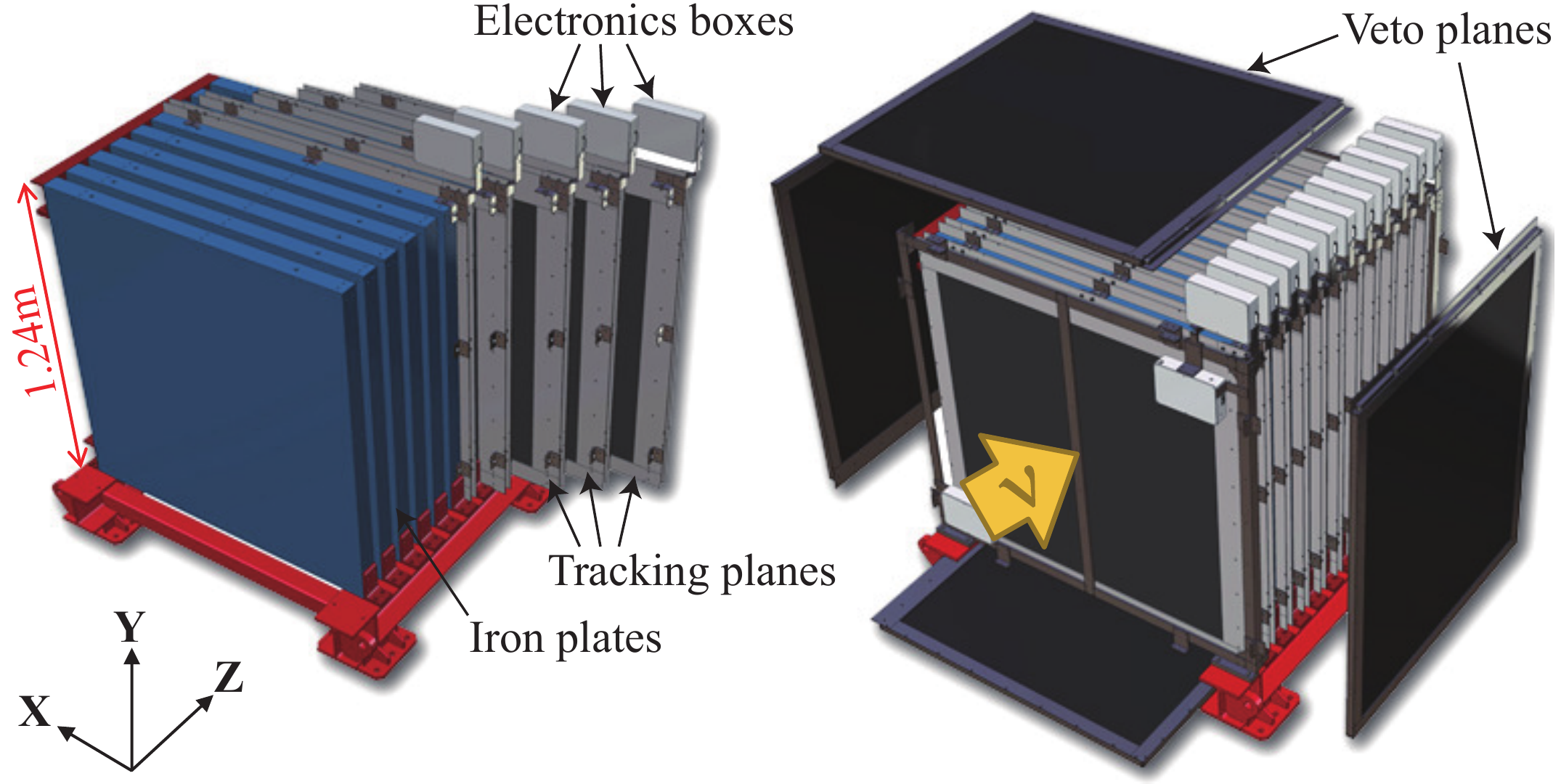}
  \caption{Exploded view of an INGRID standard module. It consists of iron target plates and tracking scintillator planes (left), and it is surrounded by veto scintillator planes (right).}
  \label{ingrid_module}
  \end{center}
\end{figure}

The main purpose of the INGRID standard modules is to monitor the neutrino beam direction with a precision better than 1 mrad.
The spatial width (1$\sigma$) of the neutrino beam at the location of INGRID is about 5~m.
In order to cover a large enough region to see a full beam profile, INGRID is designed to sample the beam in a transverse section of 10 m$\times$10 m, with 14 identical modules arranged in two identical groups along the horizontal and vertical axes, as shown in Fig.~\ref{ingrid_overview}. Two separate modules are placed off the main cross to monitor the asymmetry of the beam.
Each of the modules consists of nine iron target plates and eleven tracking scintillator planes, as shown in Fig.~\ref{ingrid_module} left.
They are surrounded by veto scintillator planes (Fig.~\ref{ingrid_module} right) to reject charged particles coming from outside of the modules. The dimensions of each iron target plate are 124$\times$124cm$^2$ in the horizontal and vertical directions and 6.5 cm along the beam direction. The total iron mass serving as a neutrino interaction target is 7.1 tons per module.
Each tracking scintillator plane consists of two scintillator layers.
Each layer has 24 scintillator bars whose dimensions are 5cm$\times$1cm$\times$120cm, making a plane of
120$\times$120cm$^2$ in the horizontal and vertical directions and 1.0 cm along the beam direction.
One layer is placed perpendicular to the other layer in a tracking scintillator plane so that it is sensitive to both horizontal and vertical positions.
The veto scintillator plane consists of one scintillator layer which is made up of 22 scintillator bars
segmented along the beam direction, in order to identify the incoming charged particles produced
by neutrino interactions in the walls of the detector hall.
Scintillation light is collected
and transported to a photodetector with a wavelength shifting fiber (WLS fiber) which is inserted in a hole at the center of the scintillator strip.
The light is read out by a Multi-Pixel Photon Counter (MPPC)
\cite{mppc_yokoyama1, mppc_yokoyama2} attached to one end of the WLS fiber.
The electrical signal from each MPPC is digitalized to integrated charge and timing information by the Trip-t front-end board (TFB) \cite{tfb}.
The integration cycle is synchronized with the neutrino beam pulse structure.
Details of the components and the basic performance of the INGRID standard modules are described in Ref \cite{ingrid_nim}.

\subsection{INGRID Proton Module}
\begin{figure}[b]
  \begin{center}
  \includegraphics[width=54mm]{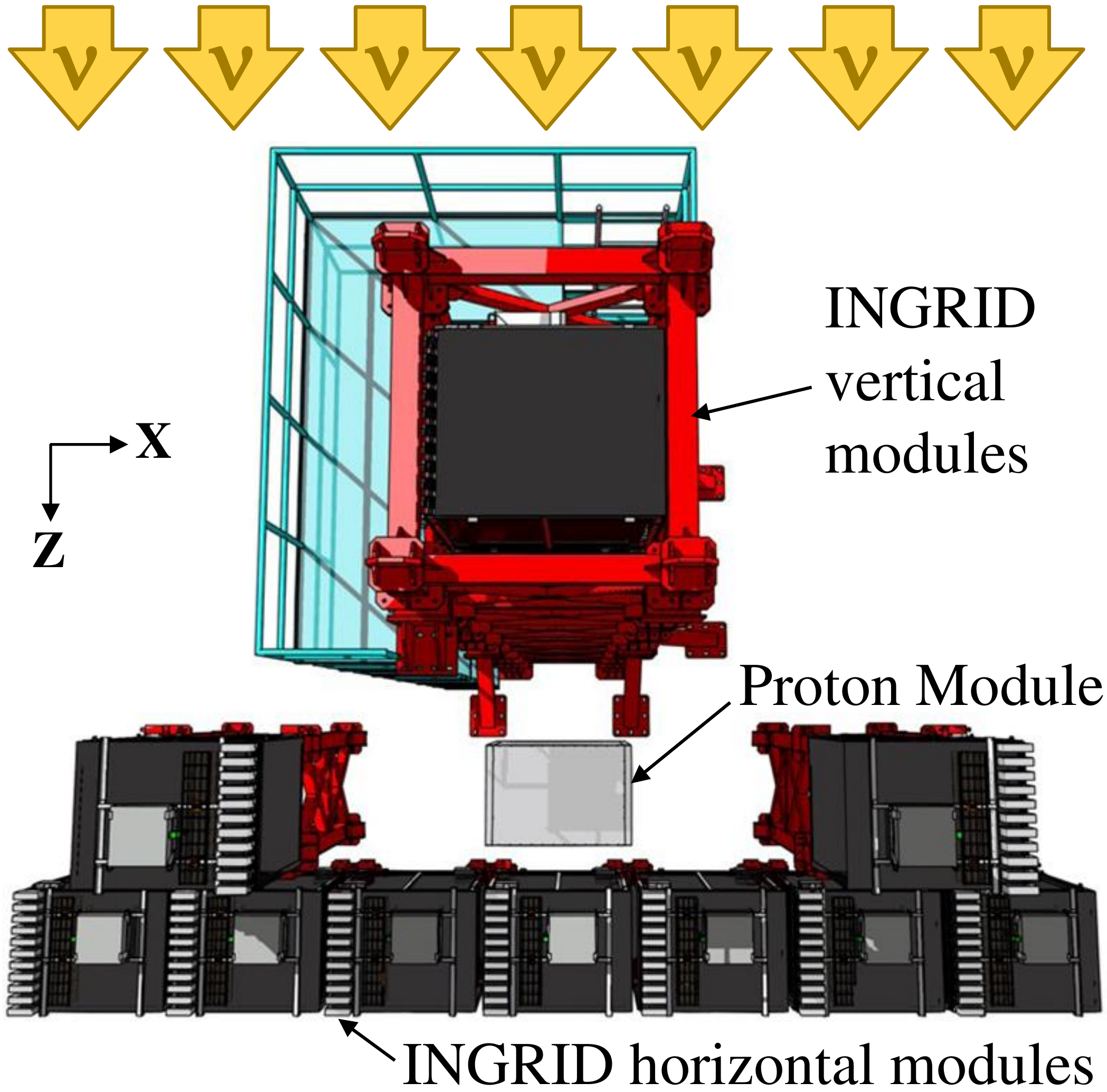}
  \caption{The position of the Proton Module viewed from above.}
  \label{pm_position}
  \end{center}
\end{figure}

The Proton Module is an extra module located at the beam center between the horizontal and vertical standard modules (Fig.~\ref{pm_position}).
It is a fully-active tracking detector which consists of only scintillator bars.
It was constructed and additionally installed between Run 1 and Run 2.
The purpose of this Proton Module is to separate the neutrino interaction channels by detecting the protons and pions together with the muons from the neutrino interactions, and to measure the neutrino cross section for each interaction channel.

\begin{figure}[htbp]
  \begin{center}
  \includegraphics[width=68mm]{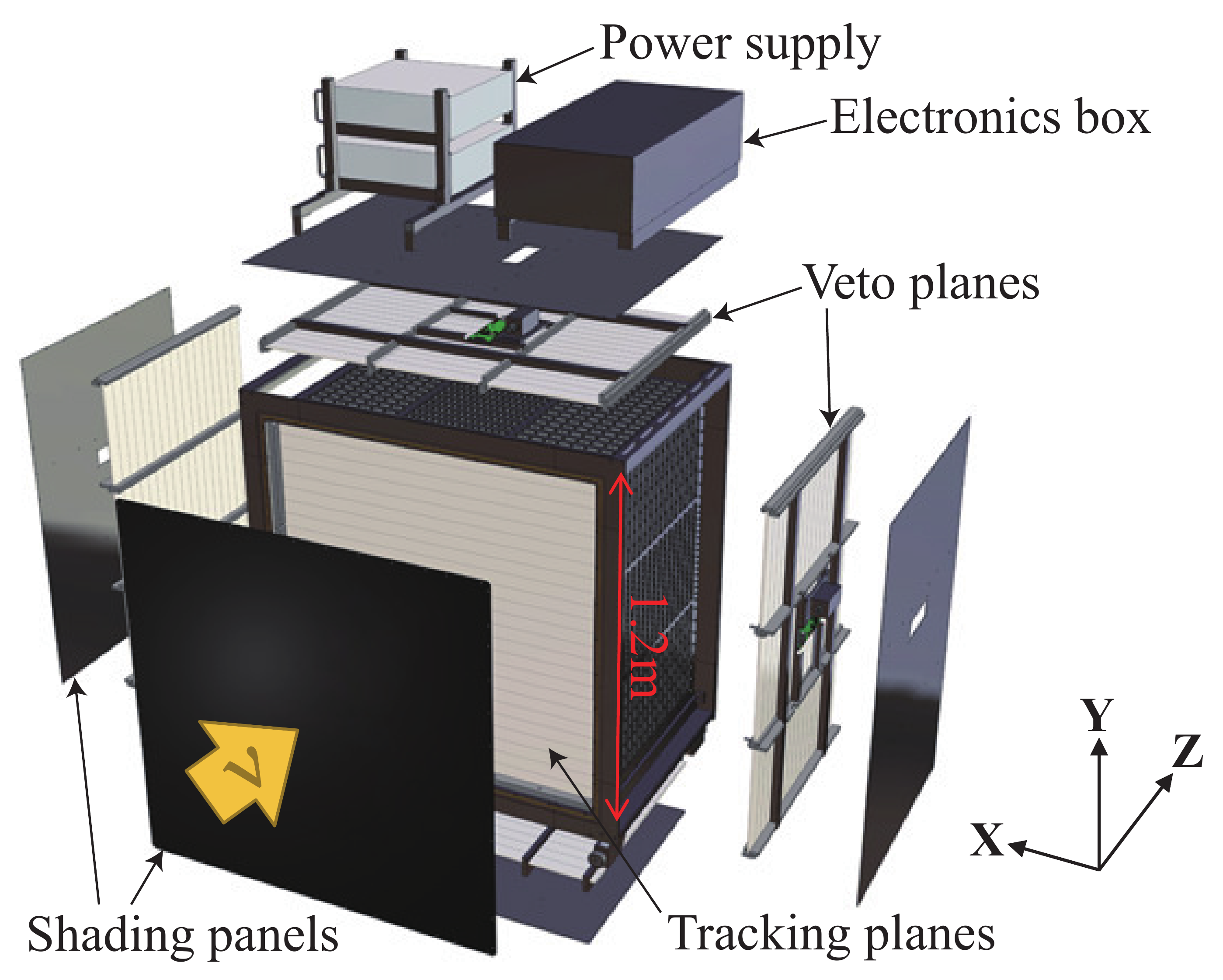}
    \caption{Exploded view of the Proton Module.}
  \label{proton_module}
  \end{center}
\end{figure}

It consists of 36 tracking layers surrounded by veto planes, where each tracking layer is an array of two types of scintillator bars (Fig.~\ref{proton_module}). The 16 bars in the inner region have dimensions of 2.5cm$\times$1.3cm$\times$120cm while the 16 bars in the outer region have dimensions of 5cm$\times$1cm$\times$120cm, making a layer of 120$\times$120cm$^2$ in the horizontal and vertical directions.
The former is the scintillator produced for the K2K SciBar detector \cite{k2k_scibar} and the latter was produced for INGRID.
The tracking layers are placed perpendicular to the beam axis at 23mm intervals. Since the bars are aligned in one direction, a tracking layer is sensitive to either the horizontal or vertical position of the tracks. The tracking layers are therefore placed alternating in perpendicular directions so that three-dimensional tracks can be reconstructed.
The tracking layers also serve as the neutrino interaction target.
As with the standard modules, scintillation light is read out by a WLS fiber and MPPC, and electrical signal from MPPC is digitalized by TFB.
The INGRID horizontal modules which lie downstream of the Proton Module are used to identify muons from the neutrino interactions in the Proton Module.

\section{Monte Carlo simulation}\label{sec:mc}
The INGRID Monte Carlo (MC) simulation consists of three main parts. The first is a simulation of the neutrino beam
production, which predicts the neutrino flux and energy spectrum of each neutrino flavor. The second is a neutrino interaction simulation, which is used to calculate the neutrino interaction cross sections and the kinematics of the final state particles taking into account the intranuclear interactions of hadrons.
The third step is a detector response simulation which reproduces the final-state particles' motion and interaction with material, scintillator light yield, and the response of the WLS fibers, MPPCs, and front-end electronics.

\subsection{Neutrino beam prediction}
\begin{figure}[htbp]
  \begin{center}
  \includegraphics[width=71mm]{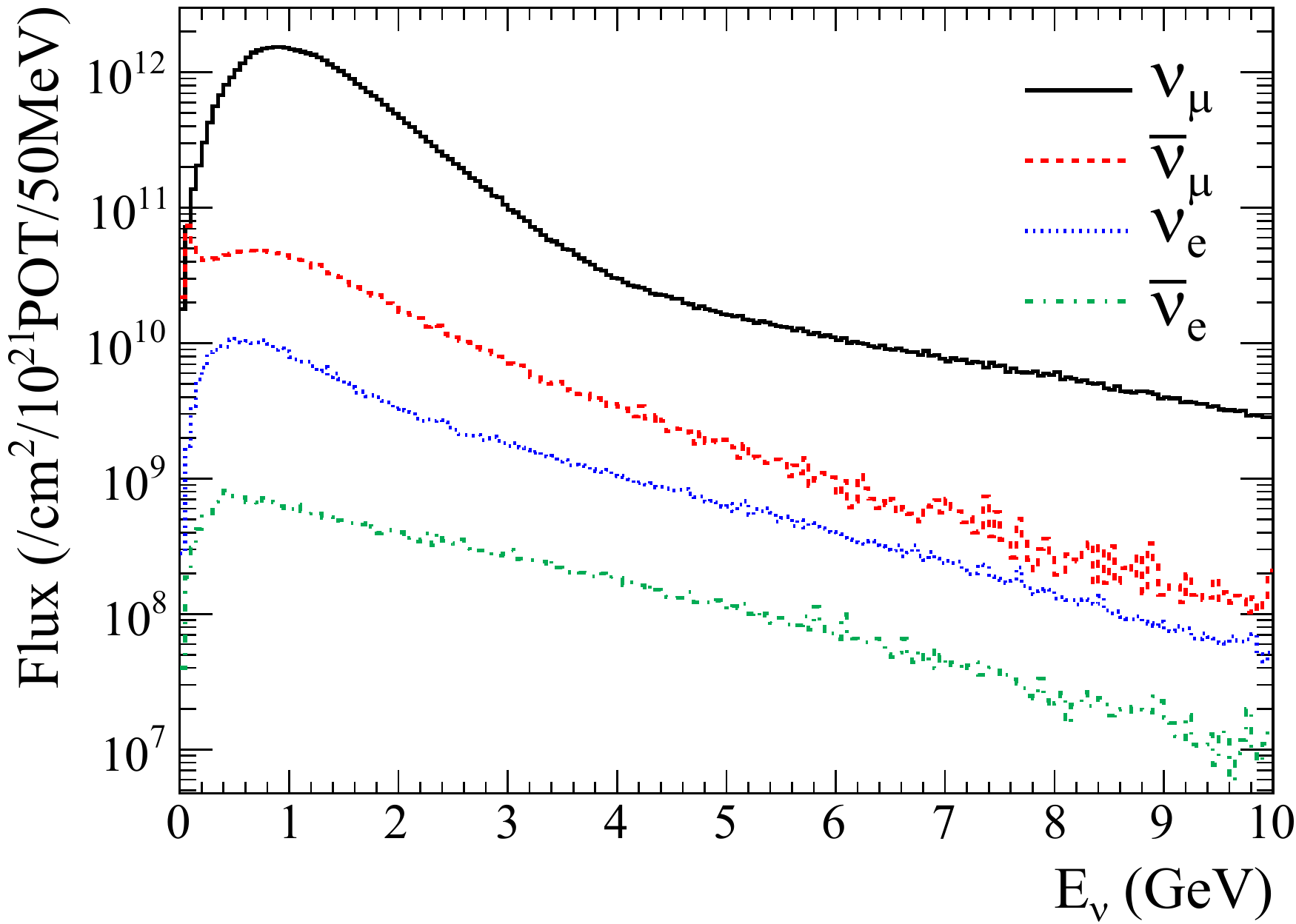}
  \caption{Neutrino energy spectrum for each neutrino species at the central module predicted by JNUBEAM.}
  \label{flux_flavor}
  \end{center}
\end{figure}

To predict the neutrino fluxes and energy spectra, a neutrino beam Monte Carlo simulation, called JNUBEAM \cite{flux_prediction}, was developed based on the GEANT3 framework \cite{geant3}. We compute the neutrino beam fluxes starting from models (FLUKA2008 \cite{fluka1,fluka2} and GCALOR \cite{gcalor}) and tune them using existing hadron production data (NA61/SHINE \cite{na61_1, na61_2}, Eichten {\it et al.} \cite{eichten} and Allaby {\it et al.} \cite{allaby}). The predicted neutrino energy spectra at the center of INGRID are shown in Fig.~\ref{flux_flavor}. Energy spectra 10 m upstream of INGRID are predicted with the same procedure in order to simulate the background events from neutrino interactions in the walls of the experimental hall.

\subsection{Neutrino interaction simulation}
\begin{figure}[htbp]
  \begin{center}
  \includegraphics[width=53mm, angle=90]{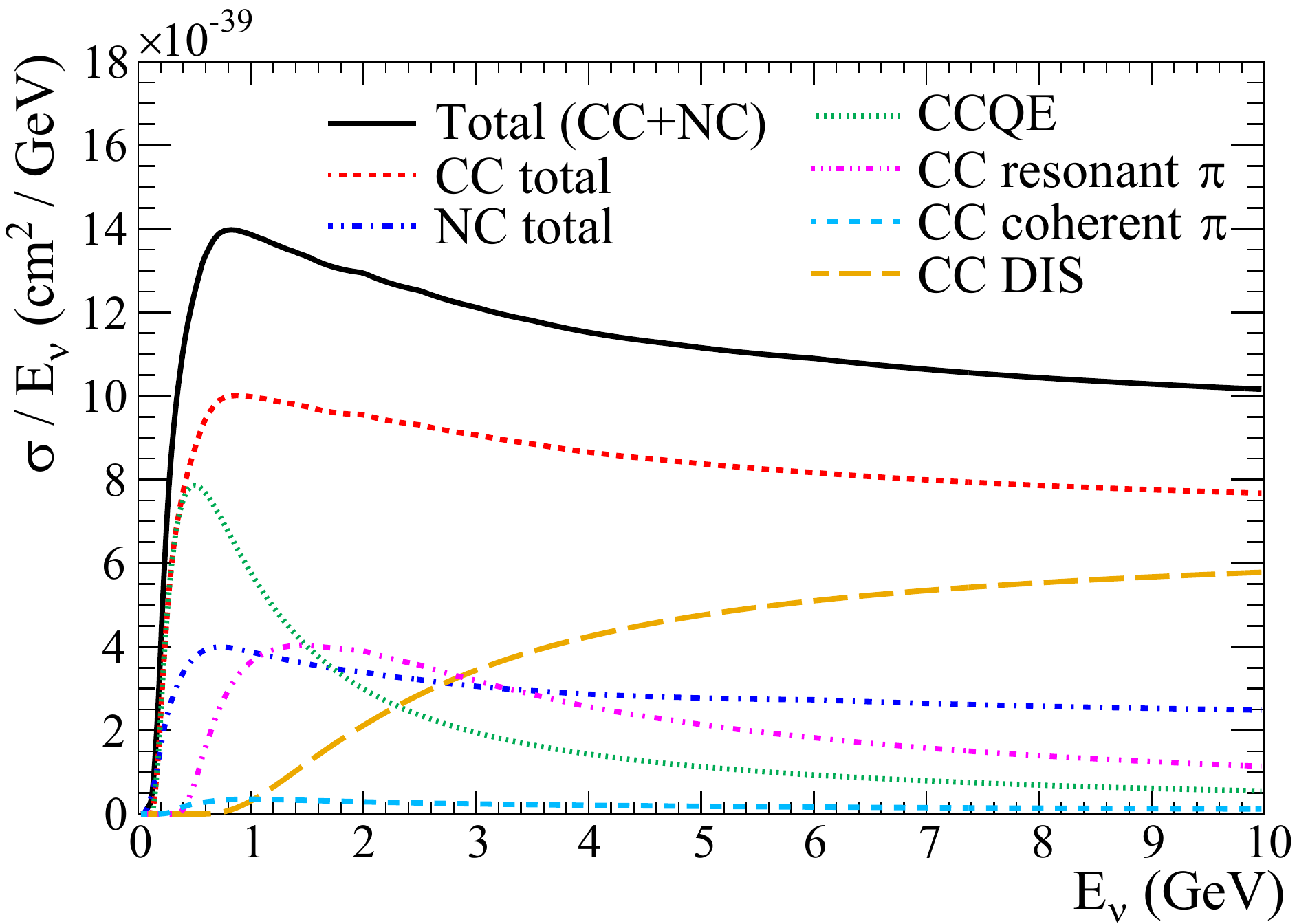}
  \caption{Neutrino-nucleus cross sections per nucleon of carbon nucleus divided by the neutrino energy predicted by NEUT.}
  \label{xsec_neut}
  \end{center}
\end{figure}

Neutrino interactions with nuclear targets are simulated with the NEUT program library \cite{neut_hayato} which has been used in the Kamiokande, Super-Kamiokande,
K2K, SciBooNE, and T2K experiments. NEUT simulates neutrino interactions with  nuclear targets 
such as protons, oxygen, carbon, and iron, in the neutrino energy range from 100~MeV to 100~TeV.
Both the primary neutrino interactions in nuclei and the secondary interactions of the hadrons in the nuclear medium are simulated.
Additionally, a cross section prediction by a different neutrino interaction simulation package, GENIE \cite{genie}, is used for comparison.
In both NEUT and GENIE, the following neutrino interactions in both charged current (CC) and neutral current (NC) are simulated:
\begin{itemize}
  \item quasi-elastic scattering ($\nu+N \rightarrow \ell+N'$)
  \item resonant $\pi$ production ($\nu+N \rightarrow \ell+\pi+N'$)
  \item coherent $\pi$ production
         ($\nu+A \rightarrow \ell+\pi\ +A'$)
  \item deep inelastic scattering ($\nu+N \rightarrow \ell+N'+\mathrm{hadrons}$)
\end{itemize}
where $N$ and $N'$ are the nucleons (proton or neutron), $\ell$ is the lepton and $A$ is the nucleus.
Both simulators use the Llewellyn Smith formalism \cite{smith}
for quasi-elastic scattering, the Rein-Sehgal model \cite{rands_res, rands_coh} for single meson production and coherent $\pi$ production and GRV98 (Gl\"{u}ck-Reya-Vogt-1998) \cite{gluck} parton distributions with Bodek-Yang modifications \cite{bodek, yang} for deep inelastic scattering.
However, the actual models used in our simulation have many differences from the above original models, such as nominal values of the axial mass, the treatment of nuclear effects, descriptions of the non-resonant inelastic scattering, etc.
For example, NEUT uses larger values of the axial mass for the quasi-elastic scattering and the resonant $\pi$ production than the world averages
based on the recent neutrino interaction measurements \cite{mb_ccqe_double_diff, mb_ccpi0, mb_ccpipm, mb_ncpi0, miniboone_ccqe}.
More details about the simulators used are described in Ref \cite{t2k_ccinc}.
Figure~\ref{xsec_neut} shows the neutrino-nucleus cross sections per nucleon divided by the neutrino energy predicted by NEUT.

\subsection{INGRID detector response simulation}
The INGRID detector simulation was developed using the Geant4 framework \cite{geant4}.
It models the real detector structures (geometries, materials).
The structure of the walls of the experimental hall is also modeled to simulate background events from neutrino interactions in the walls.
The particles' motion and physics interactions with the materials are simulated, and the energy deposit of each particle inside the scintillator is stored.
Simulations of hadronic interactions are performed with the QGSP BERT physics list \cite{qgsp_bert}.
The energy deposit is converted into the number of photons.
Quenching effects of the scintillation are modeled based on Birks' law \cite{birk1,birk2}.
The effect of collection and attenuation of the light in the scintillator and the WLS fiber is modeled based on the results of electron beam irradiation tests.
The non-linearity of the MPPC response is also taken into account, since the number of detectable photoelectrons is limited by the number of MPPC pixels. The number of photoelectrons is smeared according to statistical fluctuations and electrical noise.
The dark count of the MPPCs is added with a probability calculated from the measured dark rate.
Because the response of the ADCs on front-end electronics is not linear, its response is modeled based on the results of a charge injection test.

\section{Event selection}\label{sec:event_selection}

\subsection{Event selection for the Proton Module}
A neutrino charged current interaction in the Proton Module is identified by a track from the fiducial volume of the Proton Module to the standard horizontal modules located behind the Proton Module, where the standard modules are used to identify a long muon track.
First, hits are clustered by timing. A pre-selection is applied to reject accidental noise events. Then, tracks are reconstructed using hit information.
Next, tracks matched between the Proton Module and the standard module are searched to select long muon tracks.
If matched tracks are found, vertexing is applied to identify event pileup.
After that, charged particles from outside the module are
rejected with veto planes, and the reconstructed event vertex is required to be inside the fiducial volume.
The event selection criteria are described in the following subsections.

\subsubsection{Time clustering}
When there are four or more hits in a 100 nsec time window, all hits within $\pm$50 nsec of the average time make up a timing cluster.

\subsubsection{Pre-selection}
\begin{figure}[htbp]
  \begin{center}
  \includegraphics[width=71mm]{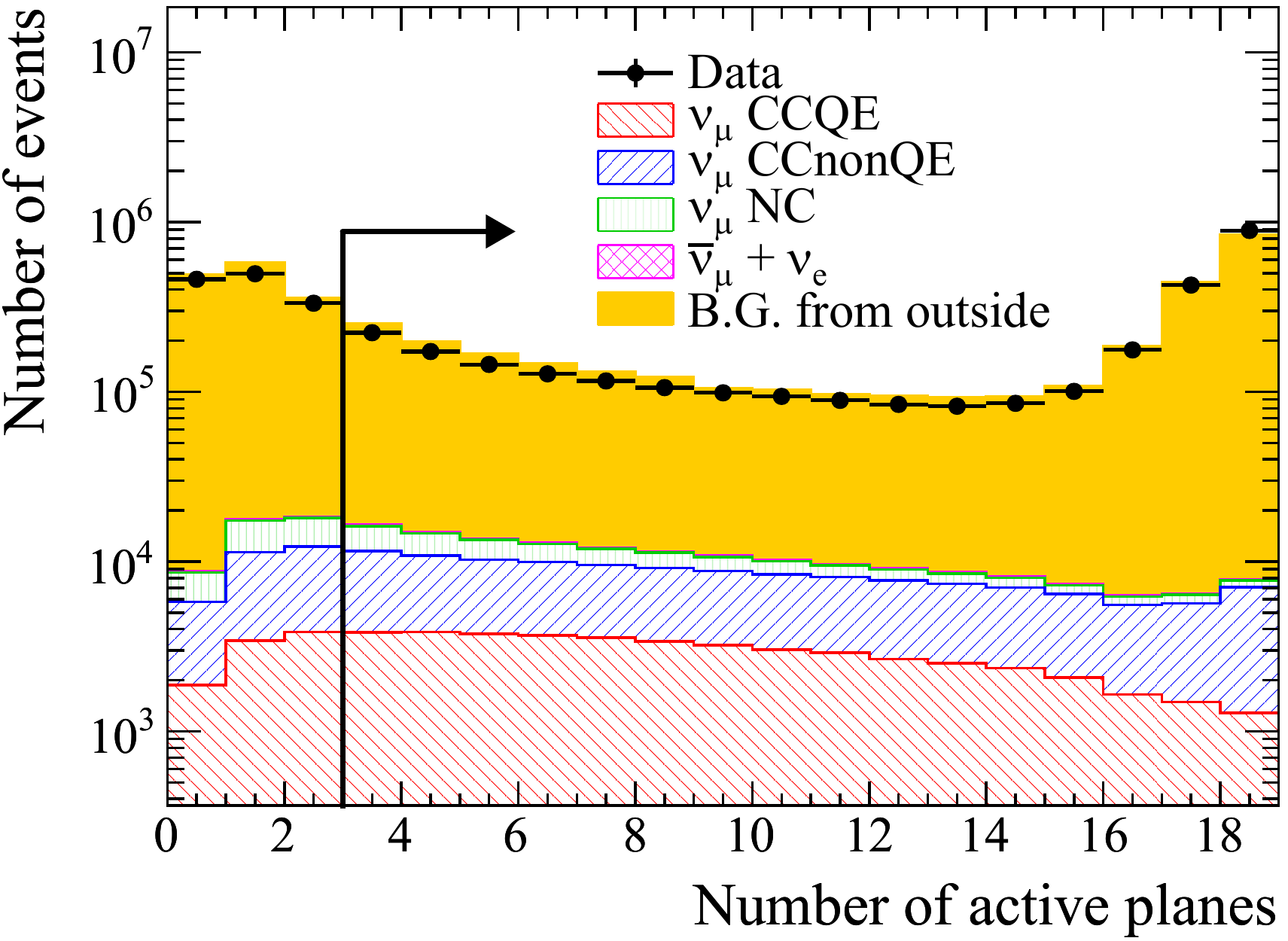}
  \caption{Number of active planes for the Proton Module. Events with more than two active planes are selected. The background events from the walls of the detector hall are normalized with beam induced muon backgrounds, as described in Sec. \ref{evt_summary_pm}.}
  \label{nactpln_pm}
  \end{center}
\end{figure}

A tracking plane with at least one hit in both the horizontal and vertical layers is defined as an active plane.
The timing clusters with three or more active planes are selected as shown in Fig.~\ref{nactpln_pm}.

\subsubsection{Two-dimensional track reconstruction}
Tracks are reconstructed in the XZ and YZ planes independently.
We developed a track reconstruction algorithm based on a cellular automaton.
The cellular automaton is the dynamical system which was used for the track reconstruction for the K2K SciBar detector \cite{cat}, and our track reconstruction algorithm is analogous with it.
This algorithm can reconstruct one or more tracks in a timing cluster.

\subsubsection{Track matching}
\begin{figure}[htbp]
  \begin{center}
  \includegraphics[width=71mm]{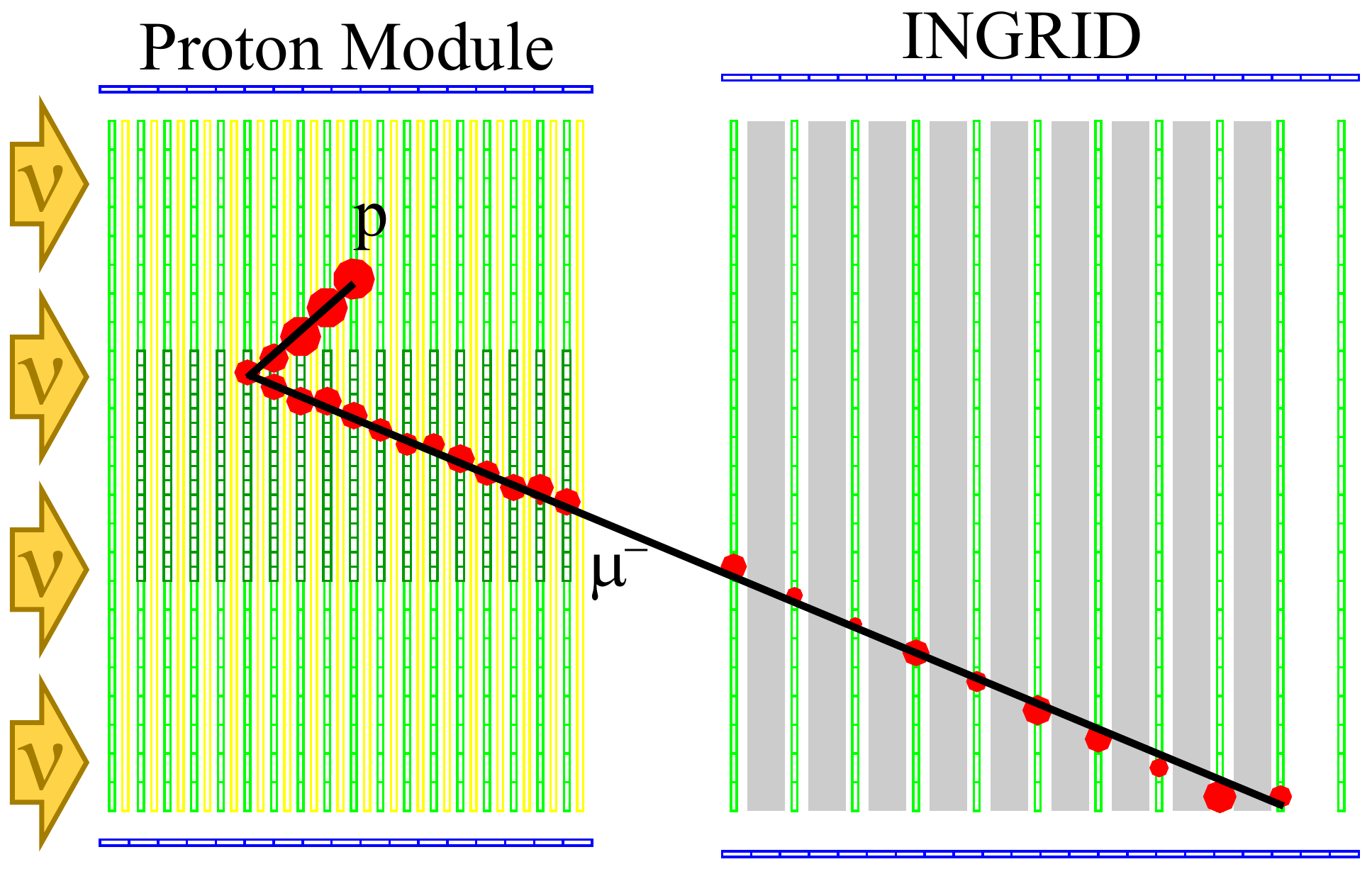}
    \caption{MC event display of a charged current neutrino event in the Proton Module. Red circles and black lines represent observed hits and reconstructed tracks, respectively. The areas of the red circles correspond to light yields.}
  \label{pm_event_display}
  \end{center}
\end{figure}

When two-dimensional tracks are reconstructed in both the horizontal standard module and the Proton Module in the same integration cycle, they are merged if they meet the following four requirements:
\begin{enumerate}
\item The upstream edge of the standard module track is in either of the most upstream two layers.
\item The downstream edge of the Proton Module track is in either of the most downstream two layers.
\item The difference between the reconstructed angles of the standard module and Proton Module tracks with respect to the z axis is less than 35$^{\circ}$.
\item At the halfway point between the standard module and the Proton Module, the distance between the extrapolated standard module track and the Proton Module track is less than 85mm.
\end{enumerate}
Figure~\ref{pm_event_display} shows an example of a merged track.
This track matching is applied to select long muon tracks from CC interactions and reject short tracks caused by neutral particles from outside, like neutrons and gammas which cannot be rejected by a veto cut, or NC interactions.

\subsubsection{Three-dimensional tracking}
Three-dimensional tracks are searched for among pairs of two-dimensional merged tracks in the XZ-plane (X track) and YZ-plane (Y track) according to the following rules.
If the difference of the upstream Z point of an X track and a Y track is smaller than three layers, they are combined into a three-dimensional track.
If a two-dimensional X or Y track meets the above condition with more than one two-dimensional Y or X track, the pair of tracks with the smallest difference in the upstream Z point is combined.

\subsubsection{Vertexing}
After the reconstruction of a three-dimensional track, the upstream edge of the three-dimensional track is identified as a reconstructed vertex.
If a pair of three-dimensional tracks meet the following conditions, they are identified as tracks coming from a common vertex:
\begin{enumerate}
\item The sum of the Z position differences between the upstream edges of the two tracks in XZ and YZ planes is less than two planes.
\item The distance between the upstream edges of the two tracks in the XY-plane is less than 150mm.
\end{enumerate}
This vertexing is performed for all combinations of three-dimensional tracks, allowing more than two tracks to be associated with the same reconstructed vertex.
The following event selection cuts are applied to every vertex, since each one is expected to correspond to a single neutrino interaction. This means that, as long as the vertices are distinguishable, events with multiple neutrino interactions (event pileup) are handled correctly.

\subsubsection{Timing cut}
\begin{figure}[htbp]
  \begin{center}
  \includegraphics[width=71mm]{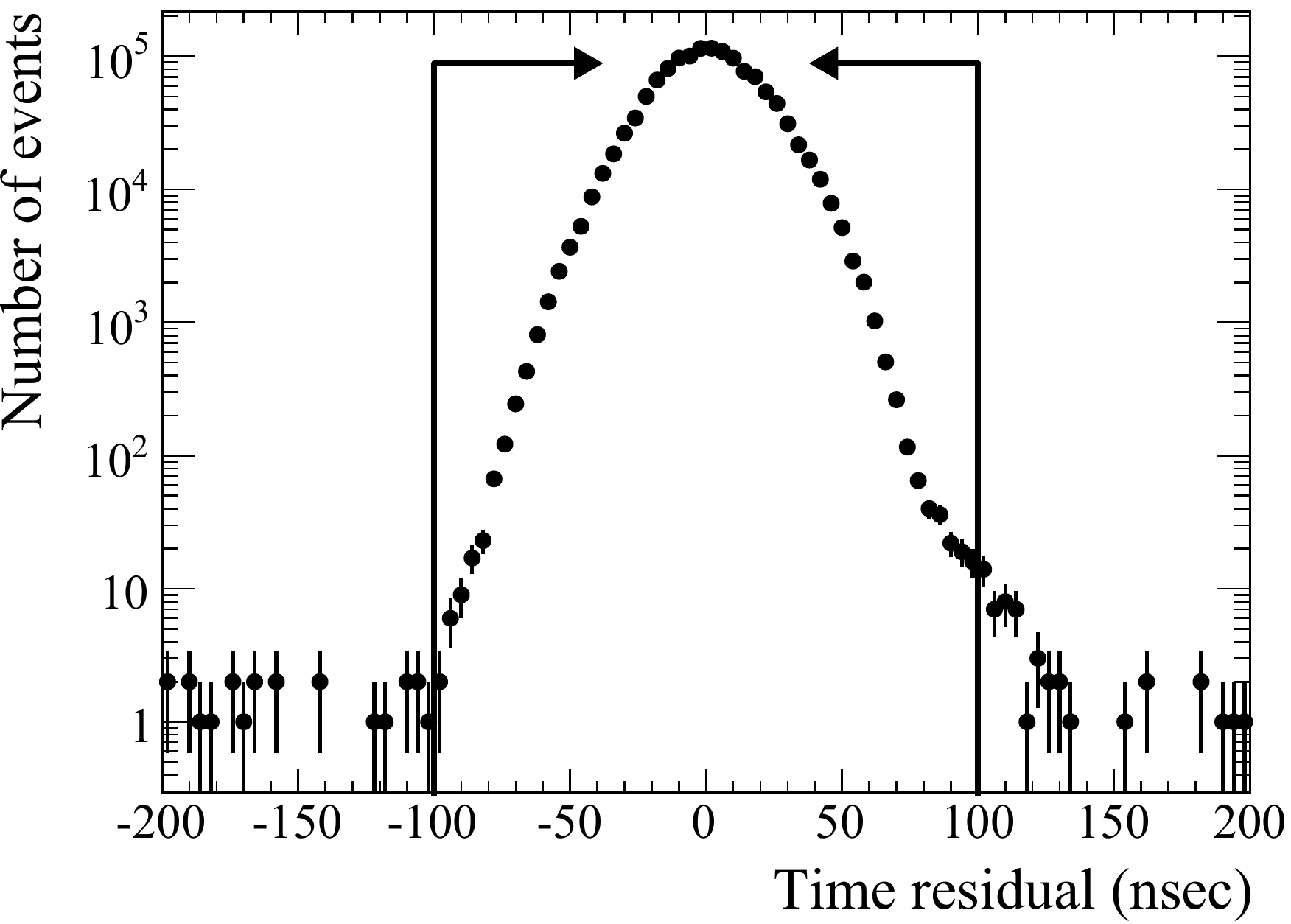}
  \caption{Time difference between the measured event timing and the expected neutrino event timing for the Proton Module. Events within $\pm$100 nsec are selected.}
  \label{timing_cut_pm}
  \end{center}
\end{figure}

The T2K neutrino beam is pulsed. Each pulse has an eight-bunch structure and each bunch has a width of 58~nsec.
To reject off-timing events, such as cosmic-ray events, only events within $\pm$100~nsec from the expected timing in each bunch are selected (Fig.\ref{timing_cut_pm}). The expected timing is calculated from the primary proton beam timing, the time of flight of the particles from the target to INGRID, and the delay of the electronics and cables. The event time is defined by the time of the hit at the start point of the track.

\subsubsection{Veto and fiducial volume cuts}
Two selections are applied to reject incoming particles produced by neutrino interactions in upstream materials, such as the walls of the experimental hall.
First, the upstream veto cut is applied.
The first tracker plane is used as the front veto plane, and events which have a vertex in the plane are rejected.
The events rejected by this front veto cut are identified as beam induced muon backgrounds.
Furthermore, events which have a hit in a side veto plane at the upstream position extrapolated from the reconstructed track are rejected.
After the veto cut, a fiducial volume (FV) cut is applied.
The FV of each module is defined as a volume within $\pm50$ cm from the module center in the X and Y directions, and from the third to the sixteenth tracker plane in the Z direction.
The ratio of the FV to the total target volume is 58.1\%.
Events having a vertex inside the FV are selected.



\subsubsection{Summary of the event selection for the Proton Module}\label{evt_summary_pm}
\begin{figure*}[htbp]
 \begin{minipage}[t]{0.325\hsize}
  \begin{center}
  \includegraphics[width=56mm]{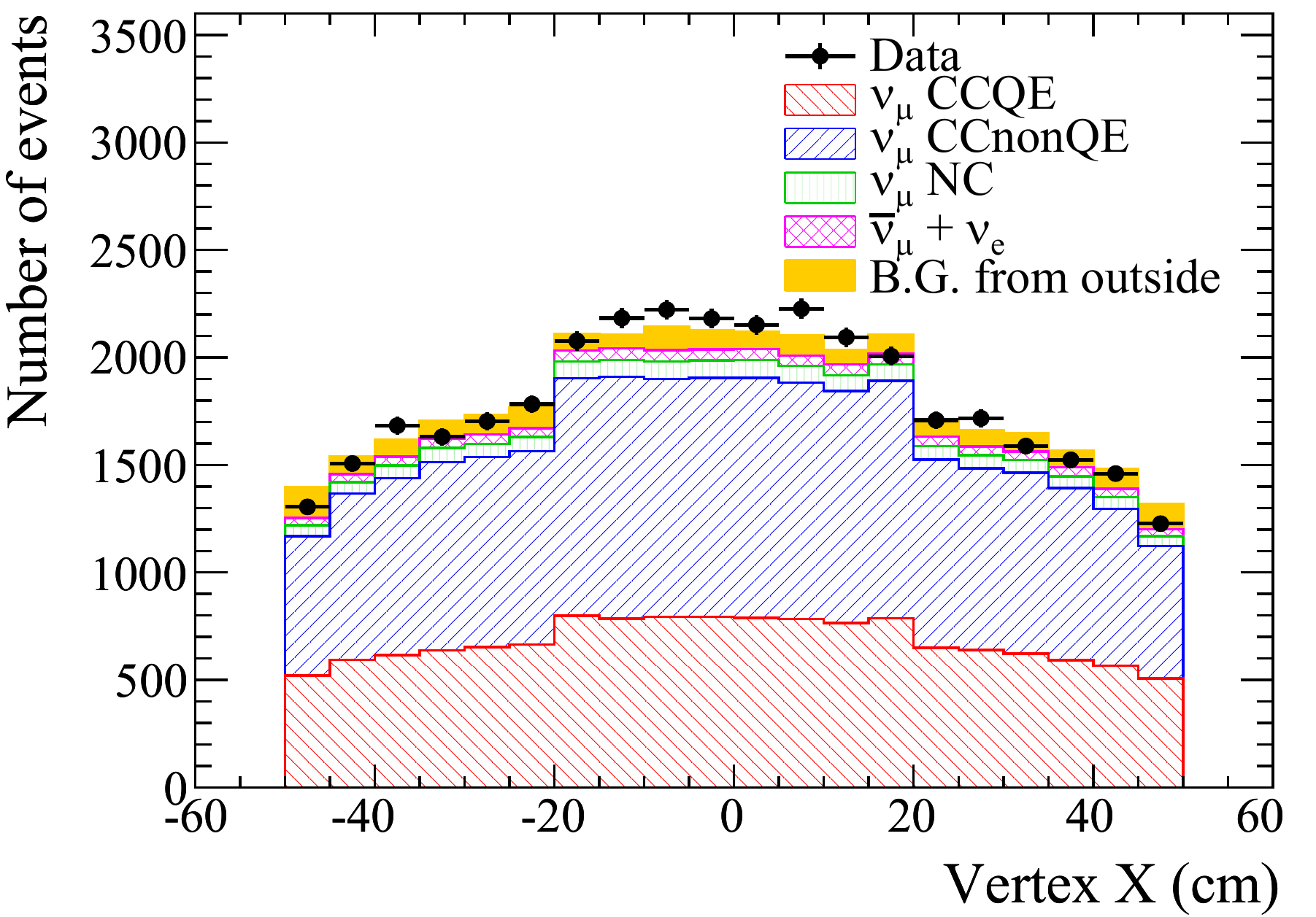}
    \end{center}
 \end{minipage}
 \begin{minipage}[t]{0.325\hsize}
  \begin{center}
  \includegraphics[width=56mm]{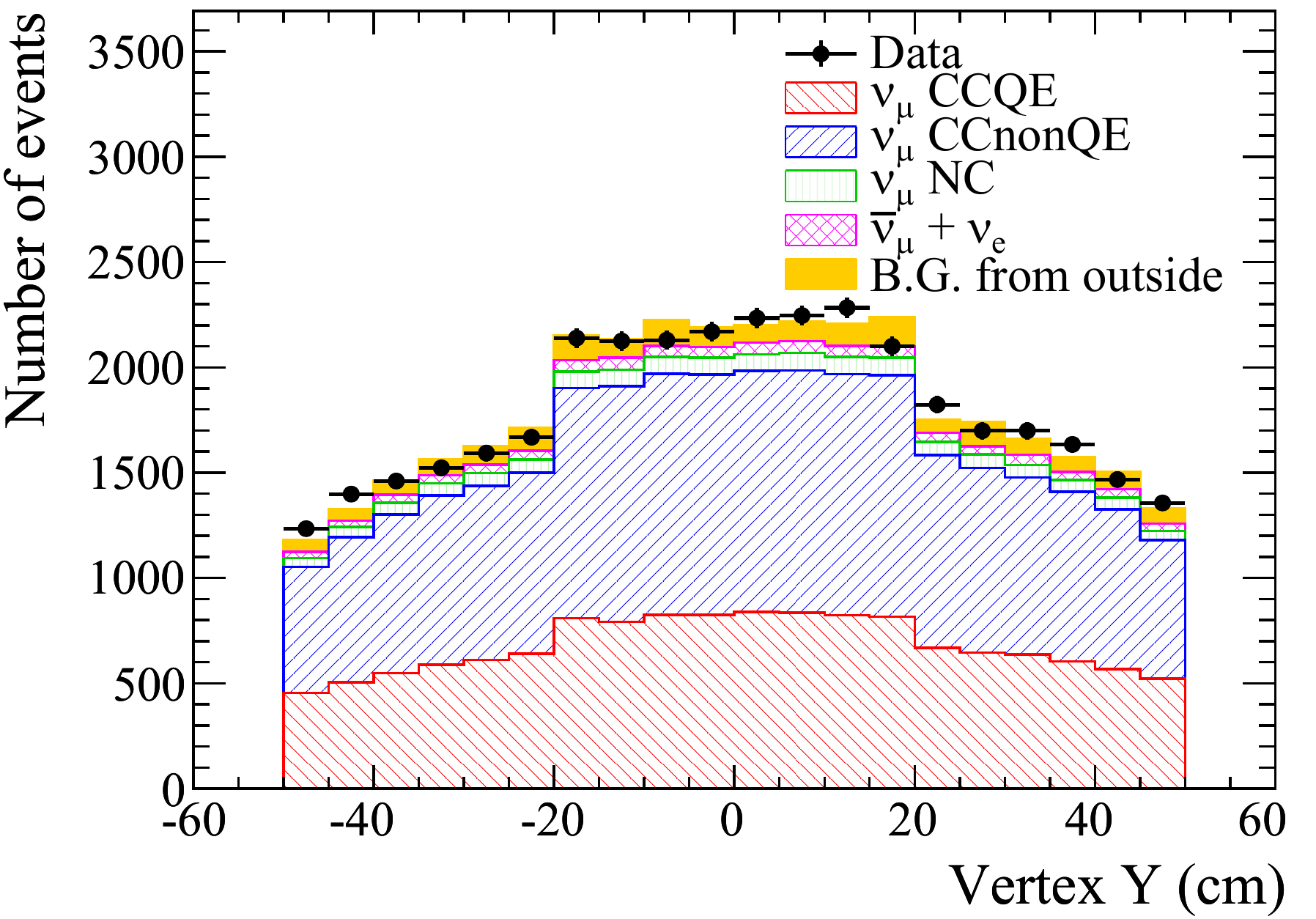}
  \end{center}
 \end{minipage}
  \begin{minipage}[t]{0.325\hsize}
  \begin{center}
  \includegraphics[width=56mm]{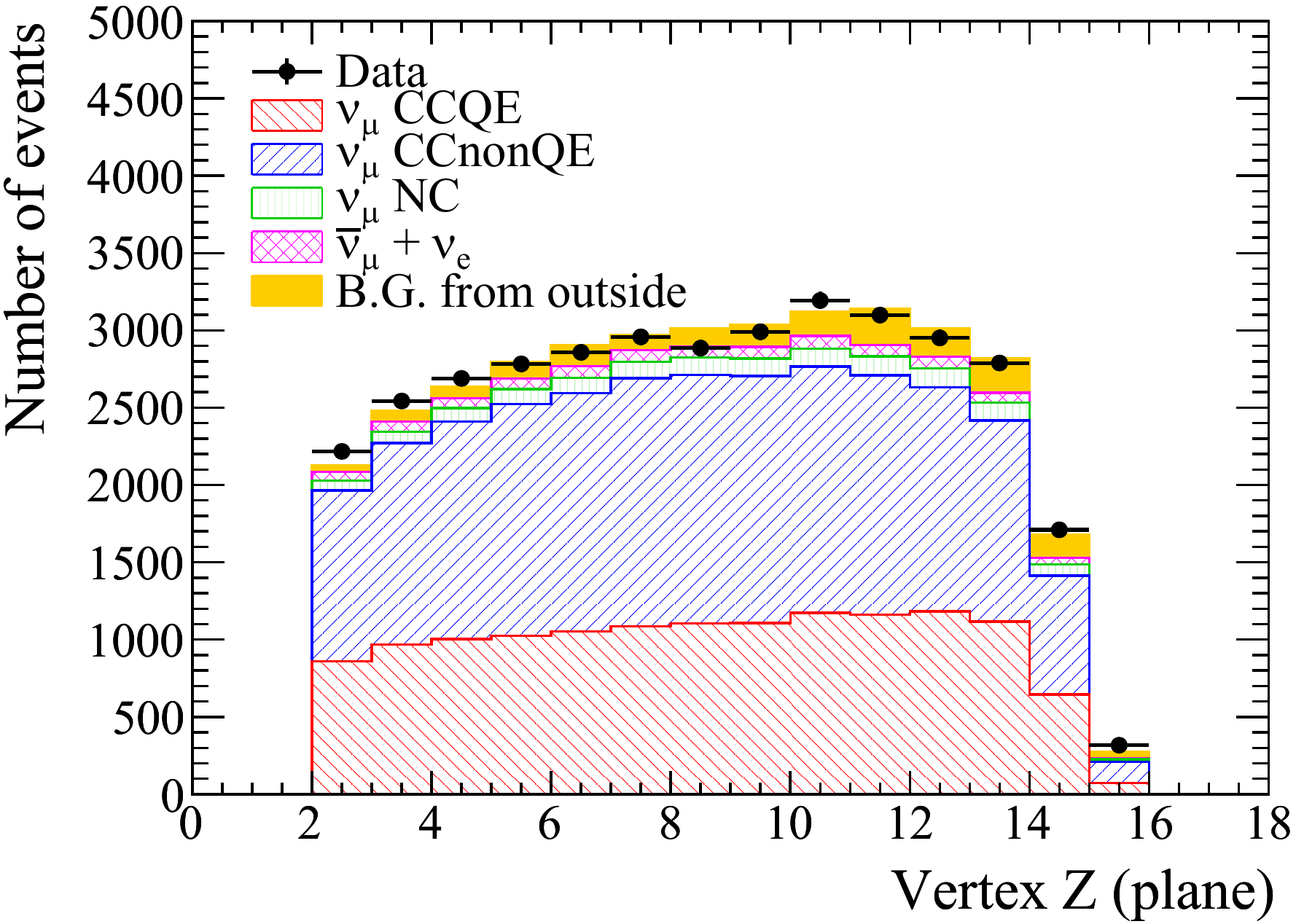}
  \end{center}
 \end{minipage}
     \caption{Vertex X, Y and Z distributions for the Proton Module following event selection. There are jumps at X or Y $=\pm20$cm because the Proton Module uses thicker scintillators in the inner region ($-20$cm $\sim+20$cm).}
  \label{vertex_pm}
\end{figure*}

The results of the event selection for the Proton Module are summarized in Table~\ref{event_selection_pm}.
Figure~\ref{vertex_pm} shows the vertex distributions in the X, Y and Z directions after all cuts.
The MC simulation includes neutrino interactions on the wall of the detector hall.
The MC prediction of the beam induced muon backgrounds is 35\% smaller than the
observation. This is likely due to the uncertainties of the density of the walls, the neutrino flux and the neutrino interaction model.
Thus, the number of neutrino interactions on the walls in the MC simulation is normalized by the observed number of the beam induced muon backgrounds.

\begin{table}[htbp]
\begin{center}
\caption{Number of events passing each selection step for the Proton Module.
The MC assumes $6.04\times10^{20}$ POT and uses the nominal NEUT model.
The efficiency is defined as the number of selected CC events divided by the number of CC interactions in the FV. The purity is defined as the fraction of the $\nu_\mu$ CC events on CH among the selected events.}
\begin{tabular}{lrrrr} \hline \hline
Selection &Data &MC &Efficiency&Purity\\ \hline
Vertexing & 1.296$\times 10^{6}$ & 1.317$\times 10^{6}$ &65.6\%&3.9\%\\
Timing cut & 1.294$\times 10^{6}$ & 1.317$\times 10^{6}$&65.6\%&3.9\%\\
Veto cut & 1.281$\times 10^{5}$ &1.380$\times 10^{5}$&53.0\%&29.9\%
\\
FV cut & 3.618$\times 10^{4}$ & 3.585$\times 10^{4}$&41.2\%&89.4\%\\
\hline \hline
\end{tabular}
\label{event_selection_pm}
  \end{center}
\end{table}

\subsection{Event selection for the standard module}
\begin{figure}[htbp]
  \begin{center}
  \includegraphics[width=71mm]{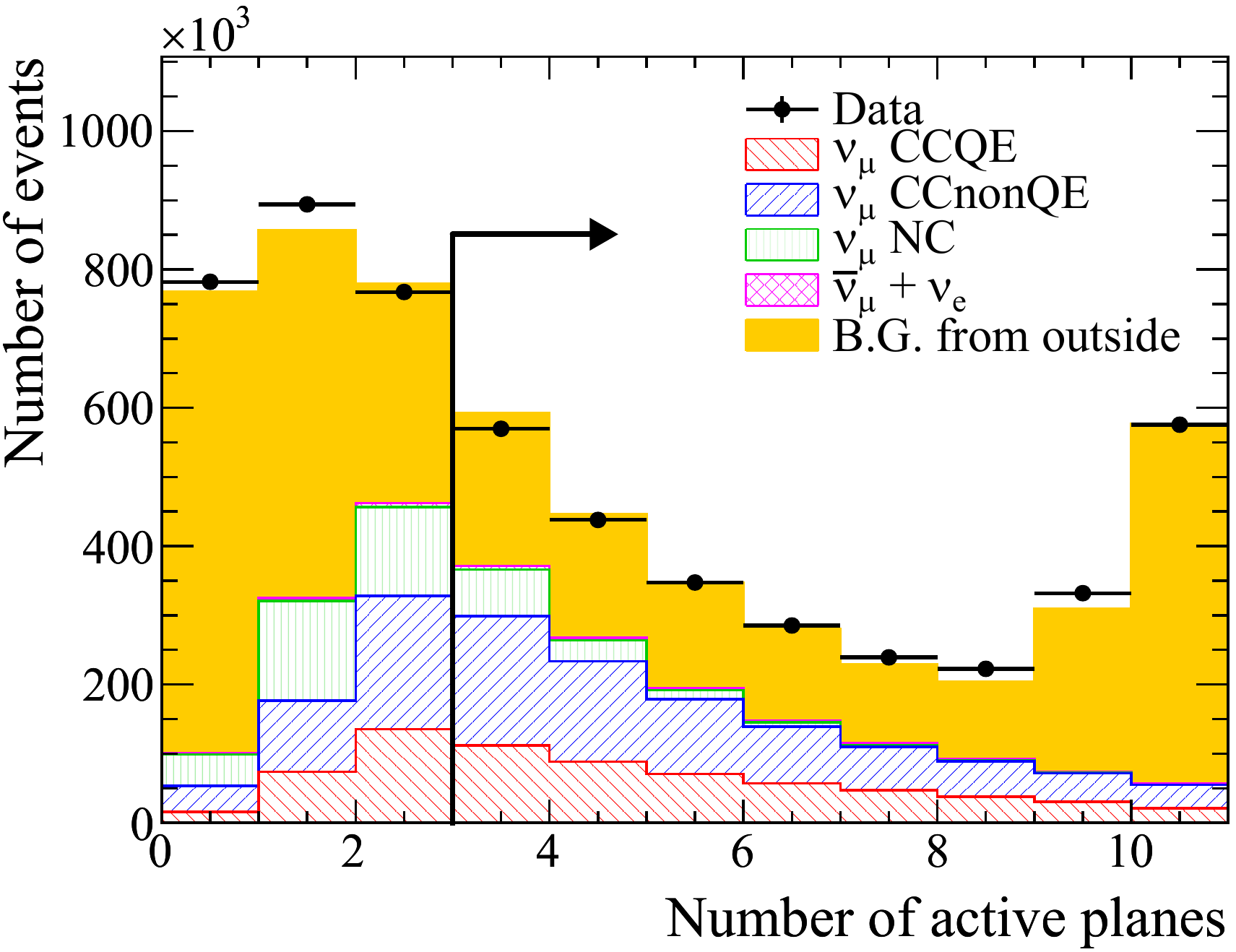}
  \caption{Number of active planes for the standard module. Events with more than two active planes are selected.}
  \label{nactpln}
  \end{center}
\end{figure}

\begin{figure}[htbp]
  \begin{center}
  \includegraphics[width=71mm]{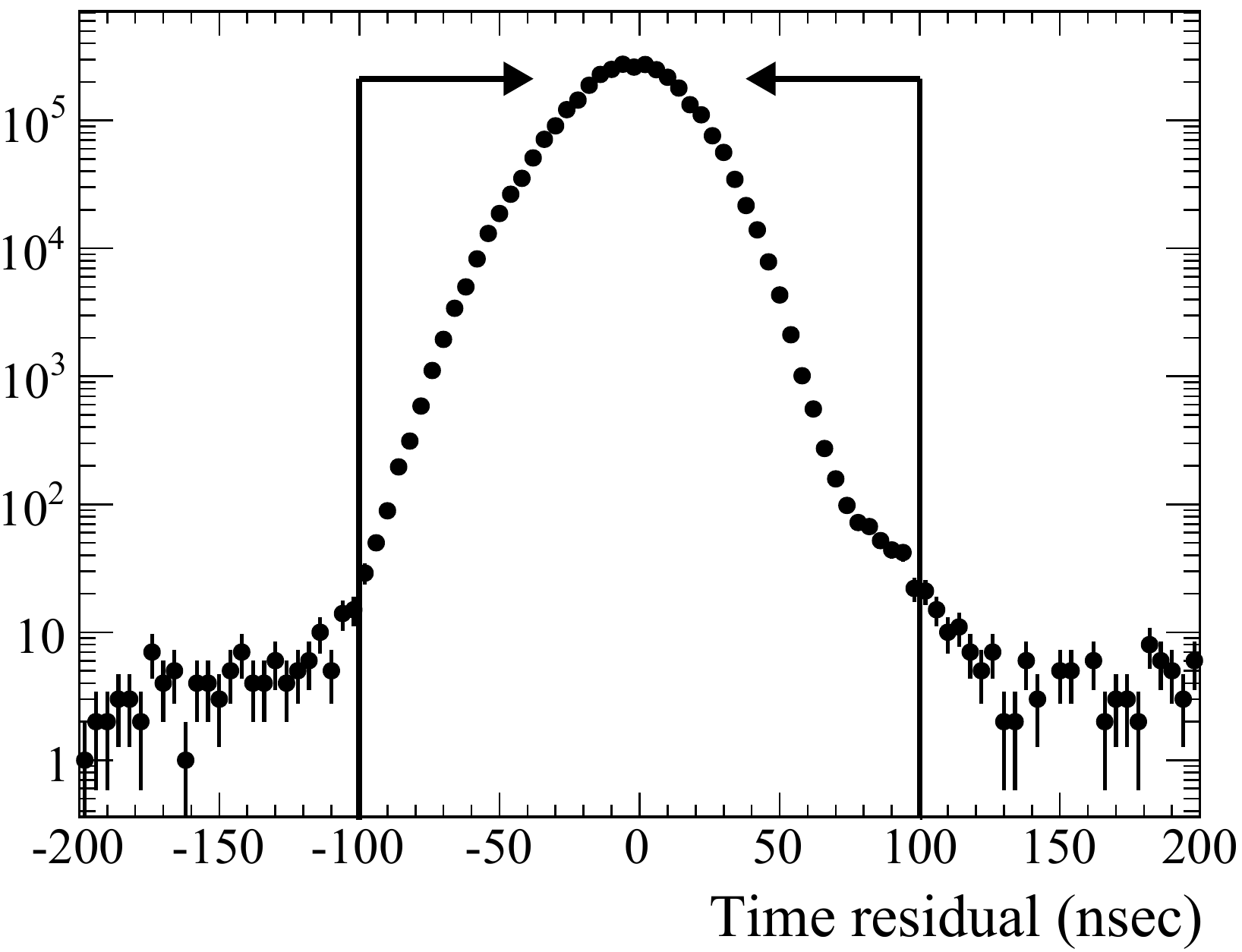}
  \caption{Time difference between the measured event timing and expected neutrino event timing for the standard module. Events within $\pm$100 nsec are selected.}
  \label{timing_cut}
  \end{center}
\end{figure}

\begin{figure*}[htbp]
 \begin{minipage}[t]{0.49\hsize}
  \begin{center}
  \includegraphics[width=71mm]{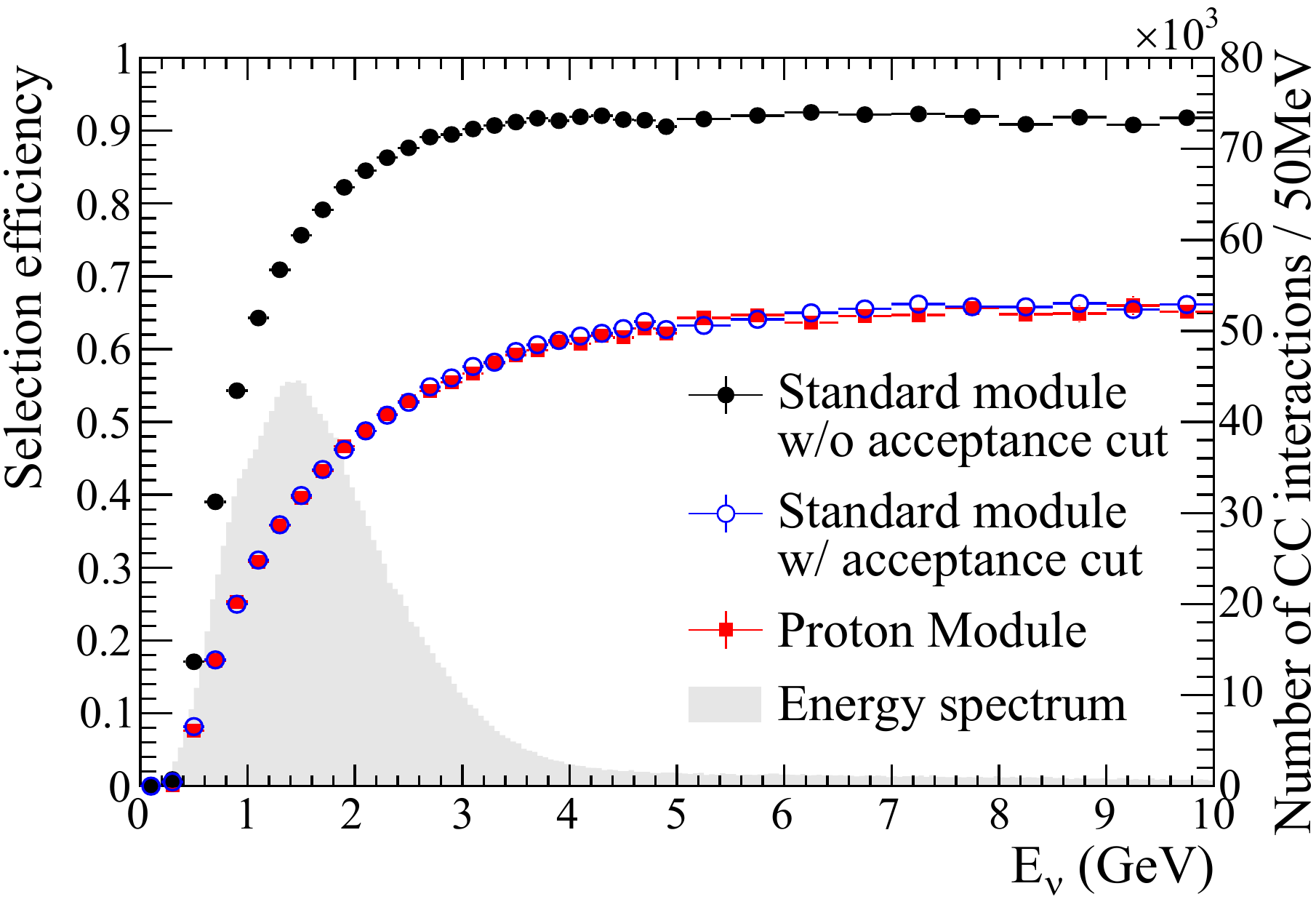}
    \end{center}
 \end{minipage}
 \begin{minipage}[t]{0.49\hsize}
  \begin{center}
  \includegraphics[width=71mm]{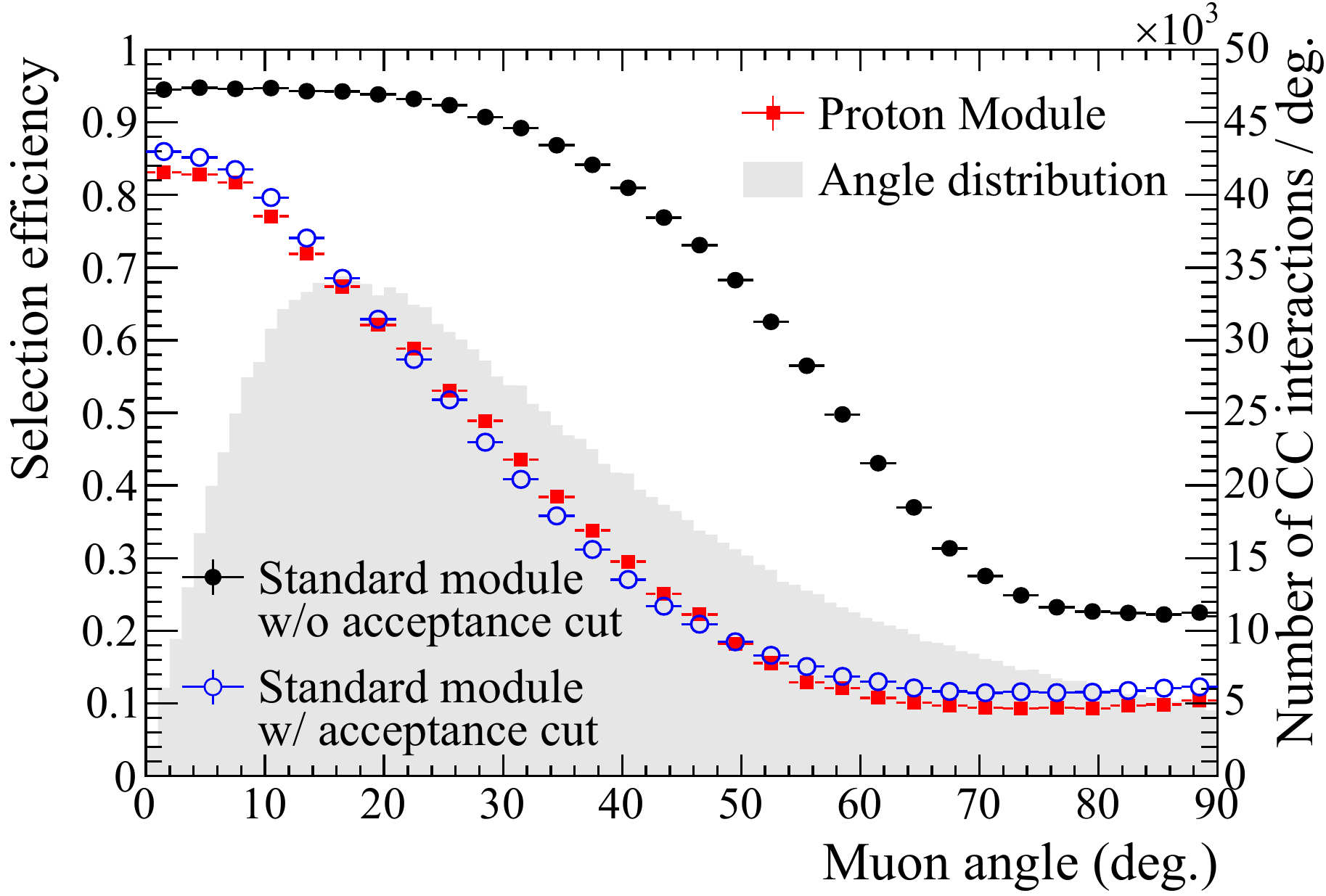}
  \end{center}
 \end{minipage}
     \caption{Event selection efficiency of CC interactions for the standard module and the Proton Module as a function of true neutrino energy (left) and true muon scattering angle (right). The energy spectrum and angle distribution of the CC interactions in the standard module are overlaid.}
  \label{eff_comparison}
\end{figure*}

\begin{figure}[htbp]
  \begin{center}
  \includegraphics[width=71mm]{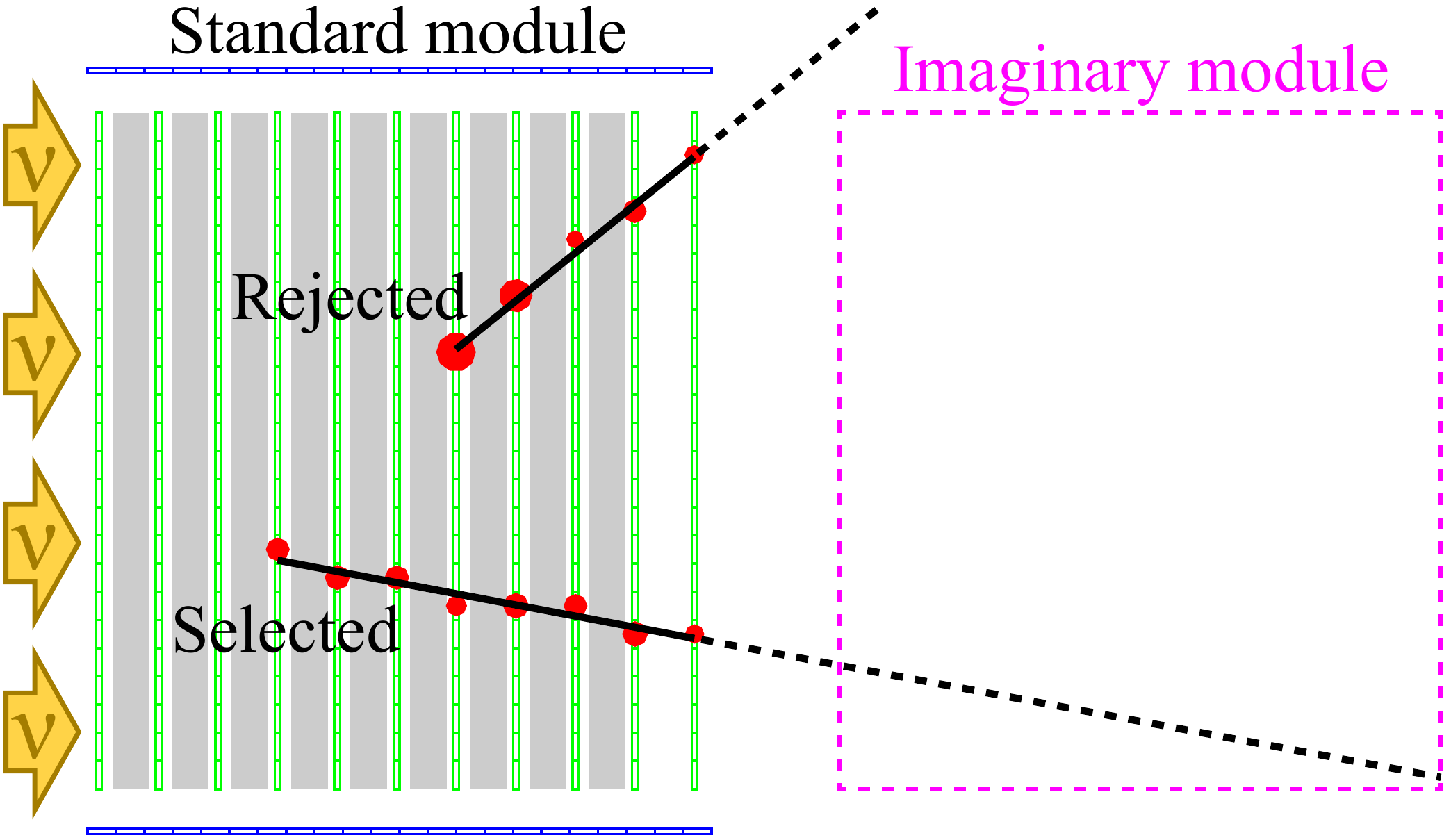}
  \caption{MC event display of a selected event and a rejected event by the acceptance cut.}
  \label{acceptance_cut}
  \end{center}
\end{figure}

\begin{figure*}[htbp]
 \begin{minipage}[t]{0.325\hsize}
  \begin{center}
  \includegraphics[width=56mm]{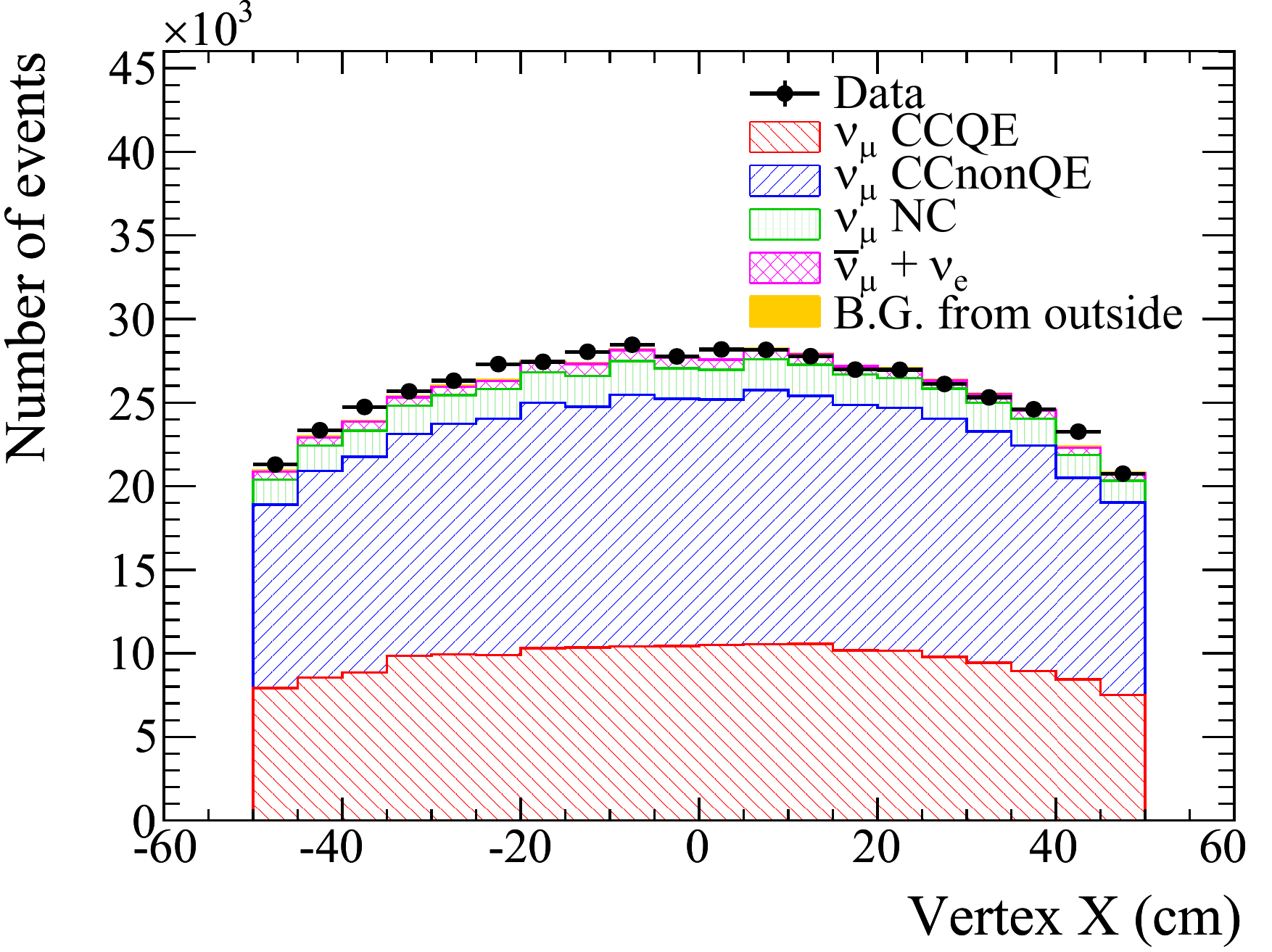}
    \end{center}
 \end{minipage}
 \begin{minipage}[t]{0.325\hsize}
  \begin{center}
  \includegraphics[width=56mm]{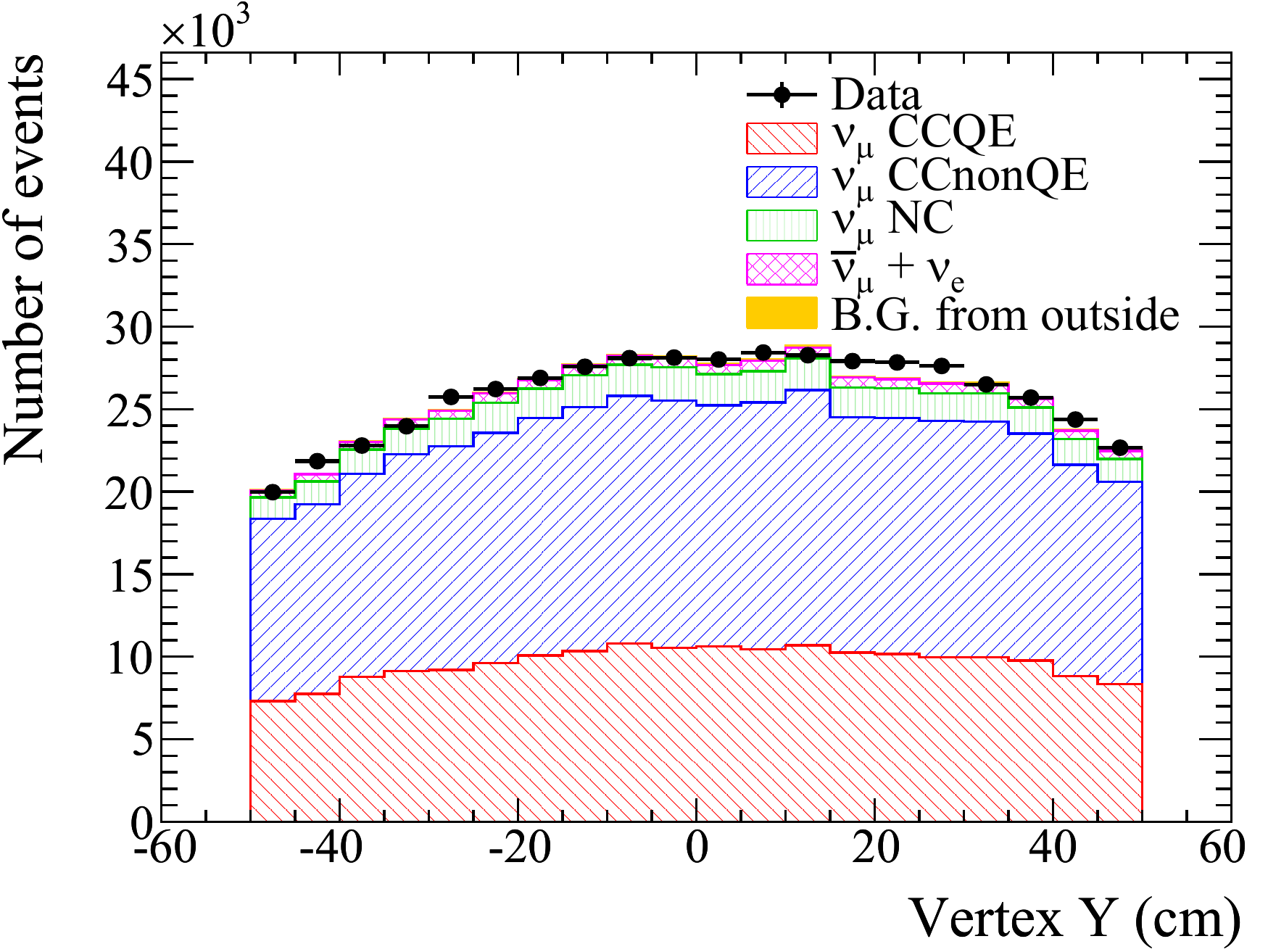}
  \end{center}
 \end{minipage}
  \begin{minipage}[t]{0.325\hsize}
  \begin{center}
  \includegraphics[width=56mm]{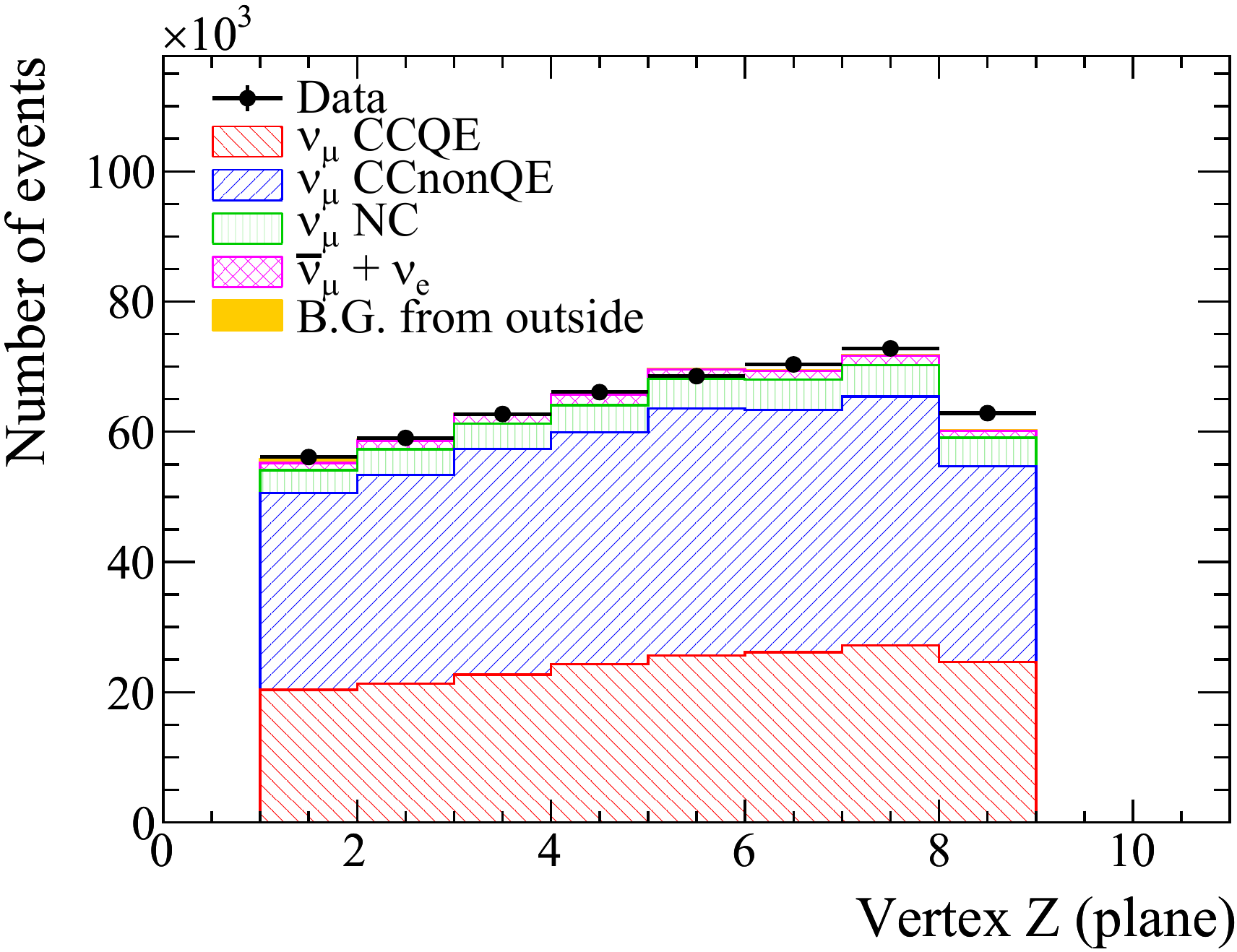}
  \end{center}
 \end{minipage}
     \caption{Vertex X, Y and Z distributions for the standard module following event selection.}
  \label{vertex_ingrid}
\end{figure*}

For the measurement of the cross section on Fe, only the horizontal central standard module was used because it is on the same beam axis as the Proton Module and thus provides a good cancellation of the systematic errors with the Proton Module.
Hence, the event selection for the standard module is applied only to the horizontal central module.
The event selection criteria for the standard module are the same as that for the Proton Module except for two differences.
One is that track matching is not applied for the standard module, and the other is an additional acceptance cut.
The event selection for the standard module is as follows.
First, time clustering, pre-selection (Fig.~\ref{nactpln}) and two-dimensional track reconstruction are applied as with the Proton Module.
When the tracks are reconstructed, three-dimensional tracking is done for all reconstructed tracks, while it is done only for the merged tracks in the case of the Proton Module.
Then, the vertexing, timing cut (Fig.~\ref{timing_cut}), and veto and fiducial volume cuts are applied as with the Proton Module.
The ratio of the FV to the total target volume is 61.7\% for the standard modules.
CC interactions in the standard module can be selected with sufficiently high purity by the above event selection.
However, there are large differences in the selection efficiency between the standard module and the Proton Module, as shown in Fig.~\ref{eff_comparison}.
This is because the acceptance of the Proton Module is limited by the required track matching with the standard module.
These differences enlarge the systematic error on the measurement of the CC-inclusive cross section ratio on Fe to CH.
To minimize this difference, the following acceptance cut is added to the event selection for the standard module.
First, an imaginary standard module is defined directly behind the standard module.
The distance between the standard module and the imaginary module is the same as that between the Proton Module and the standard module.
The reconstructed tracks are then extended further downstream, even if the track has stopped in the module.
If no tracks from the vertex reach the imaginary module, the
 event is rejected as shown in Fig.\ref{acceptance_cut}.
After applying this acceptance cut, the difference in the selection efficiencies between the standard module and the Proton Module is greatly reduced, as shown in Fig.~\ref{eff_comparison}.
The results of the event selection are summarized in Table~\ref{event_selection_sm}.
Figure~\ref{vertex_ingrid} shows the vertex distributions in the X, Y and Z directions after all cuts.
As with the Proton Module, the number of neutrino interactions on the walls in the MC simulation is normalized with the beam induced muon backgrounds.

\begin{table}[htbp]
\begin{center}
\caption{Number of events passing each selection step for the standard module.
The MC assumes $6.04\times10^{20}$ POT and uses the nominal NEUT model. The efficiency is defined as the number of selected CC events divided by the number of CC interactions in the FV. The purity is defined as the fraction of the $\nu_\mu$ CC events on Fe among the selected events.}
\begin{tabular}{lrrrr} \hline \hline
Selection &Data &MC &Efficiency&Purity\\ \hline
Vertexing & 3.179$\times 10^{6}$ & 3.194$\times 10^{6}$&96.7\%&35.9\%\\
Timing cut & 3.179$\times 10^{6}$ & 3.194$\times 10^{6}$&96.7\%&35.9\% \\
Veto cut & 1.369$\times 10^{6}$ &1.418$\times 10^{6}$&88.8\%&74.2\% \\
FV cut & 8.875$\times 10^{5}$ & 9.169$\times 10^{5}$&74.4\%&86.6\%\\
Acceptance cut & 5.185$\times 10^{5}$ & 5.130$\times 10^{5}$&42.7\%&88.8\%\\
\hline \hline
\end{tabular}
\label{event_selection_sm}
  \end{center}
\end{table}

\subsection{Event pileup correction}
When a track from a neutrino event piles up with a track from another neutrino event, vertices may fail to be reconstructed.
Because this results in the loss of events, this event-pileup effect needs to be corrected for.
The event-pileup effect is proportional to the beam intensity.
Hence, the correction factor is estimated as a linear function of the beam intensity.
The slope of the linear function is estimated from beam data as follows.
First, the beam data are categorized into sub-samples according to the beam intensity.
In each sub-sample, all hits in INGRID from two beam bunches are summed together to make one new pseudo beam bunch.
This procedure effectively doubles the beam intensity observed by INGRID.
A slope is estimated from the number of selected events in an original beam bunch and the a pseudo beam bunch for each sub-sample.
The slopes estimated from all sub-samples are consistent with each other, and the average value of this slope is used for the correction.
This event pileup correction is applied module-by-module and bunch-by-bunch using the slope and POT per bunch which corresponds to the relevant beam intensity.
The event pileup correction gives 0.85\% and 0.40\% differences in the number of selected events in the standard module and the Proton Module respectively.

\section{Analysis method}\label{sec:method}
The flux-averaged $\nu_{\mu}$ CC inclusive cross section is calculated from the number of selected events using the background subtraction and efficiency correction:
\begin{equation}
\sigma_{\mathrm{CC}} = \frac{N_{\mathrm{sel}}-N_{\mathrm{BG}}}{\phi T\varepsilon},\label{eq_cc_inc} 
\end{equation}
where $N_{\mathrm{sel}}$ is the number of selected events from real data, $N_{\mathrm{BG}}$ is the number of selected background events predicted by MC simulation, $\phi$ is the integrated $\nu_{\mu}$ flux,  $T$ is the number of target nucleons, and  $\varepsilon$ is the detection efficiency for CC events predicted by MC simulation.
The $\nu_{\mu}$ CC inclusive cross sections on Fe and CH are measured from the number of selected events in the standard module and the Proton Module, respectively.
The $\nu_{\mu}$ CC inclusive cross section ratio on Fe to CH is measured using the results from both detectors.
The background events for this analysis consist of NC events, $\bar{\nu}_{\mu}$ events, $\nu_{e}$ events, interactions on elements other than the measuring elements in the detector (Ti or O for the Proton Module, C or H for the standard module), and background events created by neutrino interactions in the material surrounding the detector.
The expected breakout of the selected events is summarized in Table \ref{breakout}.
The rate of the background events from outside for the Proton Module is much larger than that for the standard module.
It is because the number of neutrino interactions in the Proton
Module is much smaller than that in the standard module while the number of background events from outside is at a comparable level.
$N_{\mathrm{BG}}$, $\phi$, and $\varepsilon$ are estimated using MC simulation and $T$ is calculated from the target mass measured prior to detector assembly.
These quantities are summarized in Table~\ref{xsec_quantities}.

\begin{table}[htbp]
\begin{center}
\caption{Expected breakout of the selected events. The CCQE and CCnonQE events are signal events and others are background events for this measurement.}
\begin{tabular}{lrr} \hline \hline
	&	Standard module		&	Proton Module	\\
\hline
CCQE events& 35.34\%& 34.90\%\\
CCnonQE events& 51.70\%&50.53\%\\
NC events& 6.44\%& 4.19\%\\
$\bar{\nu}_{\mu}$ events&2.04\%& 2.39\%\\
$\nu_{e}$ events&0.99\%&0.73\%\\
Other target elements& 2.67\%&1.39\%\\
Backgrounds from outside&0.82\%&5.87\%\\\hline \hline
\end{tabular}
\label{breakout}
  \end{center}
\end{table}

\begin{table}[htbp]
\begin{center}
\caption{Summary of the inputs for the cross section calculation.}
\begin{tabular}{cccccc} \hline \hline
 & $N_{\mathrm{sel}}$ & $N_{\mathrm{BG}}$ & $\phi$ & $T$ & $\varepsilon$ \\ \hline
$\sigma_{\mathrm{CC}}^{\mathrm{Fe}}$ & 523045 & 67838  & 2.999$\times10^{13}$cm$^{-2}$ & 2.461$\times10^{30}$ & 0.4270\\
$\sigma_{\mathrm{CC}}^{\mathrm{CH}}$ & 36330 & 5385.5 & 3.025$\times10^{13}$cm$^{-2}$ & 1.799$\times10^{29}$  & 0.4122\\
\hline \hline
\end{tabular}
\label{xsec_quantities}
  \end{center}
\end{table}

\section{Systematic errors}\label{sec:error}
Uncertainties on $N_{\mathrm{BG}}$, $\phi$, $T$, and $\varepsilon$ are sources of systematic errors on the cross section results.
The sources of systematic error can be categorized into three groups: those from the neutrino flux prediction, the neutrino interaction model including intra-nuclear interactions, and the detector response.

\subsection{Neutrino flux uncertainties}
The neutrino flux uncertainty sources can be separated into two categories: hadron production uncertainties and T2K beamline uncertainties.
The uncertainties on hadron production are mainly
driven by the NA61/SHINE measurements \cite{na61_1, na61_2} and the
Eichten and Allaby data \cite{eichten, allaby}, and constitute the dominant
component of the flux uncertainty.
They include the uncertainties on the production cross section, the secondary nucleon production, the pion production multiplicity, and the kaon
production multiplicity.
The second category of flux uncertainties is associated
with inherent uncertainties and operational variations in the beamline conditions.
They include uncertainties in the proton beam position, the off-axis angle, the absolute horn
current, the horn angular alignment, the horn field
asymmetry, the target alignment, and the proton beam intensity.
The method of estimating these flux uncertainties is described in Ref \cite{flux_prediction}.
To evaluate the systematic error from the flux uncertainties, the flux is fluctuated using a covariance matrix in bins of the neutrino energy which is produced based on the flux uncertainties.
This is repeated for many toy data sets, and the $\pm$1$\sigma$ of the change in the cross section result is taken as the systematic error associated with the neutrino flux.

\subsection{Neutrino interaction uncertainties}
We use a data-driven method to calculate the neutrino interaction uncertainties, where the NEUT predictions are compared to available
external neutrino-nucleus data in the energy region relevant for T2K.
We fit some parameters of the models implemented in NEUT, and introduce ad hoc parameters, often with large uncertainties, to take into account remaining discrepancies between NEUT and the external data from the MiniBooNE, NOMAD, MINER$\nu$A, K2K, SciBooNE and MINOS experiments \cite{mb_ccqe_double_diff, mb_ccpi0, mb_ccpipm, mb_ncpi0, miniboone_ccqe, nomad_ccqe, minerva_ccqe_nu, minerva_ccqe_nubar, minerva_cc1pi, k2k_coh, sciboone_coh, sciboone_nccoh, minos_ccinc}.
The model parameters include axial mass values for quasi-elastic scattering and meson production via baryon resonances, the Fermi momentum, the binding energy, a spectral function parameter, and a $\pi$-less $\Delta$ decay parameter.
NEUT uses the relativistic Fermi gas model as a nuclear model.
The spectral function model is more sophisticated, and is known to be a better representation of the nuclear model.
A spectral function parameter is introduced to take into account the difference between the two nuclear models.
In the resonant pion production process, baryon resonances, mainly $\Delta$, can interact with other nucleons and disappear without pion emissions.
The $\pi$-less $\Delta$ decay parameter is introduced to take into account uncertainties on this process.
The implemented ad hoc parameters include neutrino cross section normalizations.
In addition, uncertainties on the secondary interactions of the pions with the nuclear medium are included.
Table~\ref{int_par_uncertainty} shows the nominal values and uncertainties on these parameters.
The method used to estimate these uncertainties is described in Ref \cite{t2k_ccinc}.
Systematic errors due to these parameters are estimated from variations of the cross section results when these parameters are varied within their uncertainties.
For the measurement of the CC-inclusive cross section ratio on Fe to CH, we assume that the uncertainties of
$M_A^{\mathrm{RES}}$, CC1$\pi$ normalizations, NC normalizations, Fermi gas parameters and pion secondary interactions are fully correlated between the Fe target and the CH target cases because these uncertainties are understood as independent of the target nucleus.
By contrast, the uncertainties of
$M_A^{\mathrm{QE}}$, CCQE normalizations, CC coherent pion normalization and the spectral function parameter are assumed to be uncorrelated because nuclear dependences of these uncertainties are not well understood.
In addition, the uncertainty of the CC other shape parameter which scales the number of the other CC interaction events (mainly CC deep inelastic scattering events) as a function of the neutrino energy is left out of the cross section ratio measurement because there is no evidence for a large nuclear modification in the deep inelastic scattering regime.

\begin{table}[htbp]
\begin{center}
\caption{The nominal values of and the uncertainties on the neutrino interaction parameters. The first, second, and third groups represent the model parameters, the ad hoc parameters (the neutrino cross section normalization parameters), and the scaling parameters of the pion secondary interaction probabilities.}
\begin{tabular}{ccc} \hline \hline
Parameter &  Nominal value & Error\\ \hline
$M_A^{\mathrm{QE}}$	&	1.21GeV	&	16.53\%	\\
$M_A^{\mathrm{RES}}$	&	1.21GeV	&	16.53\%	\\
$\pi$-less $\Delta$ decay &0.2 & 20\%\\
Spectral function	&	0(Off)	&	100\%	\\
Fermi momentum for Fe&	250 MeV/c	&	12\%	\\
Fermi momentum for CH&	217 MeV/c	&	13.83\%	\\
Binding energy for Fe &	33MeV	&	27.27\%	\\
Binding energy for CH &	25MeV	&	36\%	\\
\hline
CCQE norm. ($E_{\nu}<$1.5GeV)	&	1	&	11\%	\\
CCQE norm. (1.5$<E_{\nu}<$3.5GeV)	&	1	&	10\%	\\
CCQE norm. ($E_{\nu}>$3.5GeV)	&	1	&	10\%	\\
CC1$\pi$ norm.  ($E_{\nu}<$2.5GeV)	&	1	&	21\%	\\
CC1$\pi$ norm.  ($E_{\nu}>$2.5GeV)	&	1	&	21\%	\\
CC coherent $\pi$ norm.	&	1	&	100\%	\\
CC other shape	&	0(Off)	&	40\%\\
NC1$\pi^0$ norm.	&	1	&	31\%	\\
NC coherent $\pi$ norm.	&	1	&	30\%	\\
NC1$\pi^{\pm}$ norm.	&	1	&	30\%	\\
NC other norm.	&	1	&	30\%	\\
\hline
Pion absorption	&	1	&	50\%	\\
Pion charge exchange (low energy)	&	1	&	50\%	\\
Pion charge exchange (high energy)	&	1	&	30\%	\\
Pion QE scattering (low energy)	&	1	&	50\%	\\
Pion QE scattering (high energy)	&	1	&	30\%	\\
Pion inelastic scattering	&	1	&	50\%	\\
\hline \hline
\end{tabular}
\label{int_par_uncertainty}
  \end{center}
\end{table}

\subsection{Detector response uncertainties}
The uncertainty of the target mass measurement,  0.13\% for the standard module and 0.25\% for Proton Module, is taken as the systematic error on the target mass.
Variation of the measured MPPC dark rate during data acquisition, 5.84 hits/cycle for the standard module and 11.52 hits/cycle for the Proton Module, is taken as the uncertainty on the MPPC dark rate, where the cycle denotes the integration cycle synchronized with the neutrino beam pulse structure.
The discrepancy between the hit detection efficiency measured with beam induced muon backgrounds and that of the MC simulation, 0.17\% for the standard module and 0.21\% for the Proton Module, is assigned as the uncertainty in the hit detection efficiency.
The relations between these quantities and the cross section results are estimated by MC simulation, and variations on the calculated cross section results due to these uncertainties are assigned as systematic errors.
The event pileup correction factor has uncertainties which come from the statistics of the beam data and the MPPC dark count in the estimation of the correction factor.
The systematic error from these uncertainties is estimated assuming the highest beam intensity achieved in beam operation so far.
There is about a 35\% discrepancy between the beam induced muon background rate estimated by the MC simulation and that measured from the data.
The change in the background contamination fraction from this discrepancy is taken as the systematic error for the beam-related background.
The cosmic-ray background was found to be very small from the out-of-beam timing data.
The systematic error on the track reconstruction efficiency is estimated by comparing the efficiency for several sub-samples between the data and the MC simulation.
The standard deviation of the data $-$ MC of the track reconstruction efficiency for the sub-samples is taken as the systematic error.
The systematic errors from all event selections are evaluated by varying each selection threshold.
The maximum difference between the data and MC for each selection threshold is taken as the value of each systematic error.

\subsection{Summary of the systematic errors}
Table~\ref{ccinc_syst} summarizes the systematic errors on each measurement.
The total systematic errors on the measurements of the CC inclusive cross section on Fe, that on CH, and their ratio are $_{-10.84\%}^{+13.11\%}$, $_{-10.69\%}^{+12.91\%}$, and $_{-3.33\%}^{+3.32\%}$, respectively.
The neutrino flux error is the dominant systematic error for the measurement of the CC inclusive cross section on Fe and CH.
However, it is small for the measurement of the cross section ratio on Fe to CH, since this error mostly cancels between two detectors, as expected.

\begin{table*}[htbp]
\begin{center}
\caption{Summary of the systematic errors.}
\begin{tabular}{cccc} \hline \hline
Item &  $\sigma_{\mathrm{CC}}^{\mathrm{Fe}}$ & $\sigma_{\mathrm{CC}}^{\mathrm{CH}}$ & $\sigma_{\mathrm{CC}}^{\mathrm{Fe}}/\sigma_{\mathrm{CC}}^{\mathrm{CH}}$\\ \hline
Neutrino flux	&$-$10.34\%+12.74\% &$-$10.12\%+12.48\% &$-$0.31\%+0.31\% \\
\hline
$M_A^{\mathrm{QE}}$&$-$1.44\%+1.42\%&$-$0.60\%+0.72\%&$-$1.61\%+1.55\%\\
$M_A^{\mathrm{RES}}$&$-$0.35\%+0.20\%&$-$0.61\%+0.45\%&$-$0.25\%+0.27\%\\
CCQE normalization ($E_{\nu}<$1.5GeV)&$-$0.82\%+0.79\%&$-$0.52\%+0.50\%&$-$0.95\%+0.94\%\\
CCQE normalization (1.5$<E_{\nu}<$3.5GeV)&$-$0.45\%+0.50\%&$-$0.67\%+0.76\%&$-$0.88\%+0.83\%\\
CCQE normalization ($E_{\nu}>$3.5GeV)&$-$0.11\%+0.11\%&$-$0.10\%+0.11\%&$-$0.15\%+0.15\%\\
CC1$\pi$ normalization  ($E_{\nu}<$2.5GeV)&$-$1.50\%+1.37\%&$-$1.72\%+1.66\%&$-$0.28\%+0.22\%\\
CC1$\pi$ normalization  ($E_{\nu}>$2.5GeV)&$-$0.50\%+0.52\%&$-$0.54\%+0.56\%&$-$0.04\%+0.04\%\\
CC coherent $\pi$ normalization&$-$0.48\%+0.49\%&$-$1.03\%+1.10\%&$-$1.20\%+1.14\%\\
CC other shape&$-$0.82\%+0.77\%&$-$1.07\%+1.02\%&$-$\\
NC1$\pi^0$ normalization&$-$0.30\%+0.31\%&$-$0.18\%+0.18\%&$-$0.13\%+0.13\%\\
NC coherent $\pi$ normalization&$-$0.02\%+0.02\%&$-$0.01\%+0.01\%&$-$0.01\%+0.01\%\\
NC1$\pi^{\pm}$ normalization&$-$0.31\%+0.31\%&$-$0.23\%+0.23\%&$-$0.07\%+0.07\%\\
NC other normalization&$-$1.21\%+1.23\%&$-$0.71\%+0.72\%&$-$0.51\%+0.51\%\\
$\pi$-less $\Delta$ decay&$-$0.50\%+0.54\%&$-$0.35\%+0.39\%&$-$0.15\%+0.15\%\\
Spectral function&$-$0.76\%+0.00\%&$-$0.98\%+0.00\%&$-$0.76\%+0.98\%\\
Fermi momentum&$-$0.43\%+0.49\%&$-$0.39\%+0.41\%&$-$0.04\%+0.08\%\\
Binding energy&$-$0.31\%+0.32\%&$-$0.22\%+0.25\%&$-$0.09\%+0.07\%\\
Pion absorption&$-$0.15\%+0.13\%&$-$0.09\%+0.08\%&$-$0.05\%+0.04\%\\
Pion charge exchange (low energy)&$-$0.06\%+0.09\%&$-$0.07\%+0.10\%&$-$0.16\%+0.17\%\\
Pion charge exchange (high energy)&$-$0.09\%+0.08\%&$-$0.08\%+0.08\%&$-$0.02\%+0.00\%\\
Pion QE scattering (low energy)&$-$0.14\%+0.15\%&$-$0.18\%+0.13\%&$-$0.00\%+0.06\%\\
Pion QE scattering (high energy)&$-$0.16\%+0.11\%&$-$0.23\%+0.21\%&$-$0.10\%+0.08\%\\
Pion inelastic scattering&$-$0.24\%+0.20\%&$-$0.26\%+0.23\%&$-$0.03\%+0.02\%\\
\hline
Target mass	&							$\pm$0.14\%	&	$\pm$0.27\%	&$\pm$0.30\%\\
MPPC dark count	&							$\pm$0.23\%	&	$\pm$0.12\%	&$\pm$0.26\%\\
Hit efficiency	&				$\pm$0.44\%	&	$\pm$0.44\%	&$\pm$0.62\%\\
Event pileup	&				$\pm$0.05\%	&	$\pm$0.03\%	&$\pm$0.06\%\\
Beam-related background	&		$\pm$0.10\%	&	$\pm$0.93\%	&$\pm$0.94\%\\
Cosmic-ray background	&	$\pm$0.01\%	&	$\pm$0.02\%	&$\pm$0.02\%\\
2D track reconstruction	&		$\pm$0.50\%	&	$\pm$0.58\%	&$\pm$0.77\%\\
Track matching	&	$-$		&	$\pm$0.31\%	&$\pm$0.31\%\\
3D tracking	&			$\pm$0.15\%	&	$\pm$0.97\%	&$\pm$0.98\%\\
Vertexing	&					$\pm$0.31\%	&	$\pm$0.12\%	&$\pm$0.33\%\\
Beam timing cut	&				$\pm$0.01\%	&	$\pm$0.01\%	&$\pm$0.01\%\\
Veto cut	&					$\pm$0.53\%	&	$\pm$0.58\%	&$\pm$0.79\%\\
Fiducial volume cut	&			$\pm$0.40\%	&	$\pm$0.18\%	&$\pm$0.44\%\\
Acceptance cut	&				$\pm$0.36\%	&	$-$	&$\pm$0.36\%\\ \hline
Total	&	$-$10.84\%+13.11\% & $-$10.69\%+12.91\% & $-$3.33\%+3.32\%\\
\hline \hline
\end{tabular}
\label{ccinc_syst}
  \end{center}
\end{table*}

\section{Results}\label{sec:result}
\begin{figure*}[htbp]
 \begin{minipage}[t]{0.49\hsize}
  \begin{center}
    \includegraphics[height=82.5mm,angle=90]{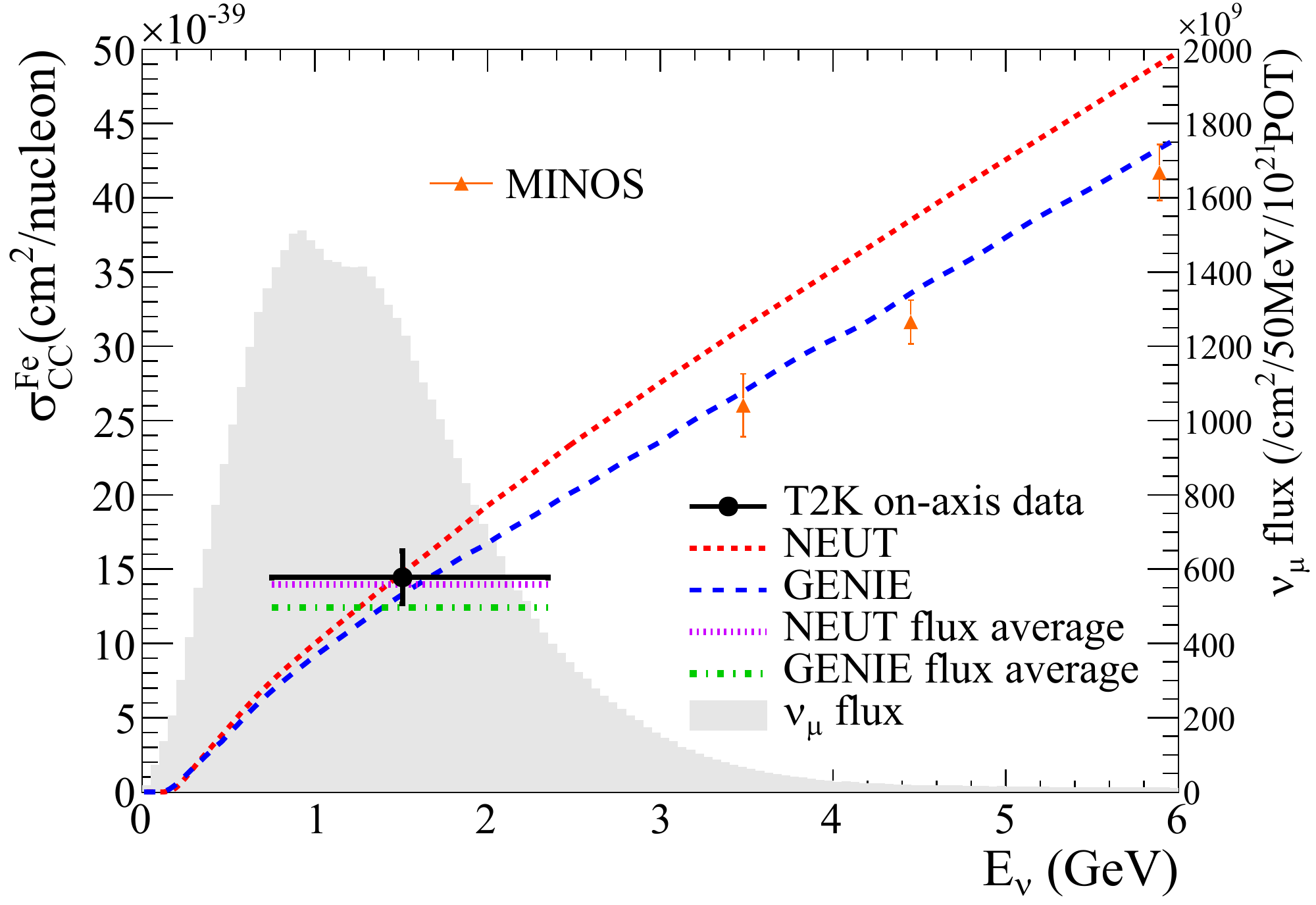}
    \end{center}
 \end{minipage}
 \begin{minipage}[t]{0.49\hsize}
  \begin{center}
      \includegraphics[height=82.5mm,angle=90]{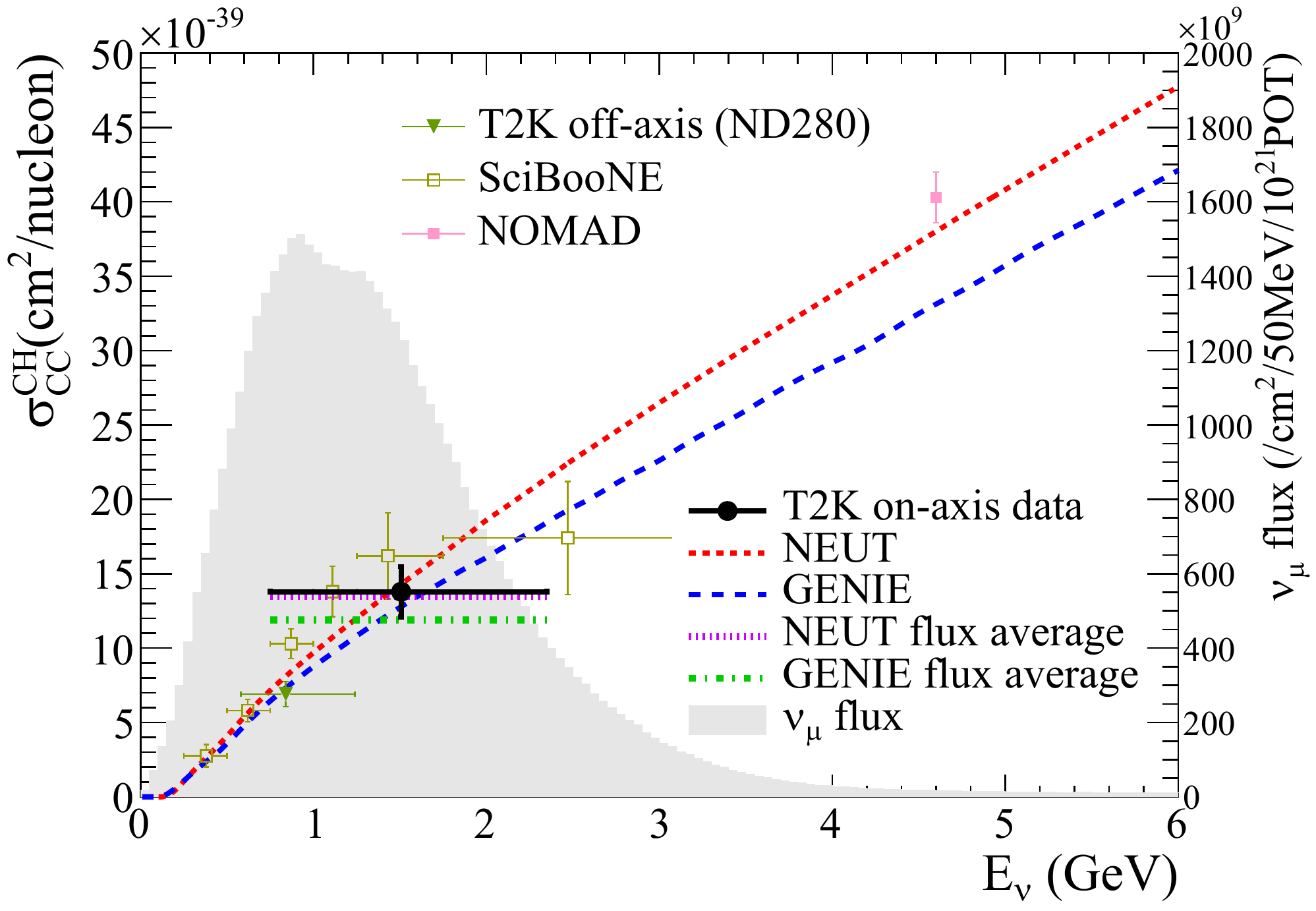}
  \end{center}
 \end{minipage}
     \caption{The inclusive $\nu_{\mu}$ charged current cross section on Fe (left) and that on CH (right) with predictions by NEUT and GENIE. The isoscalar corrections are not applied to our data or predictions. Our data point is placed at the flux mean energy. The vertical error bar represents the total (statistical and systematic) uncertainty,
and the horizontal bar represents 68\% of the flux at each side of the mean energy. The MINOS, T2K ND280, SciBooNE and NOMAD results are also plotted\cite{minos_ccinc, t2k_ccinc, sciboone_ccinc, nomad_ccinc}.
Because the isoscalar correction is applied to the MINOS data, it is expected to be shifted by about $-$2\%.}
  \label{cc_incl_result}
\end{figure*}

\begin{figure}[htbp]
  \begin{center}
  \includegraphics[height=82.5mm,angle=90]{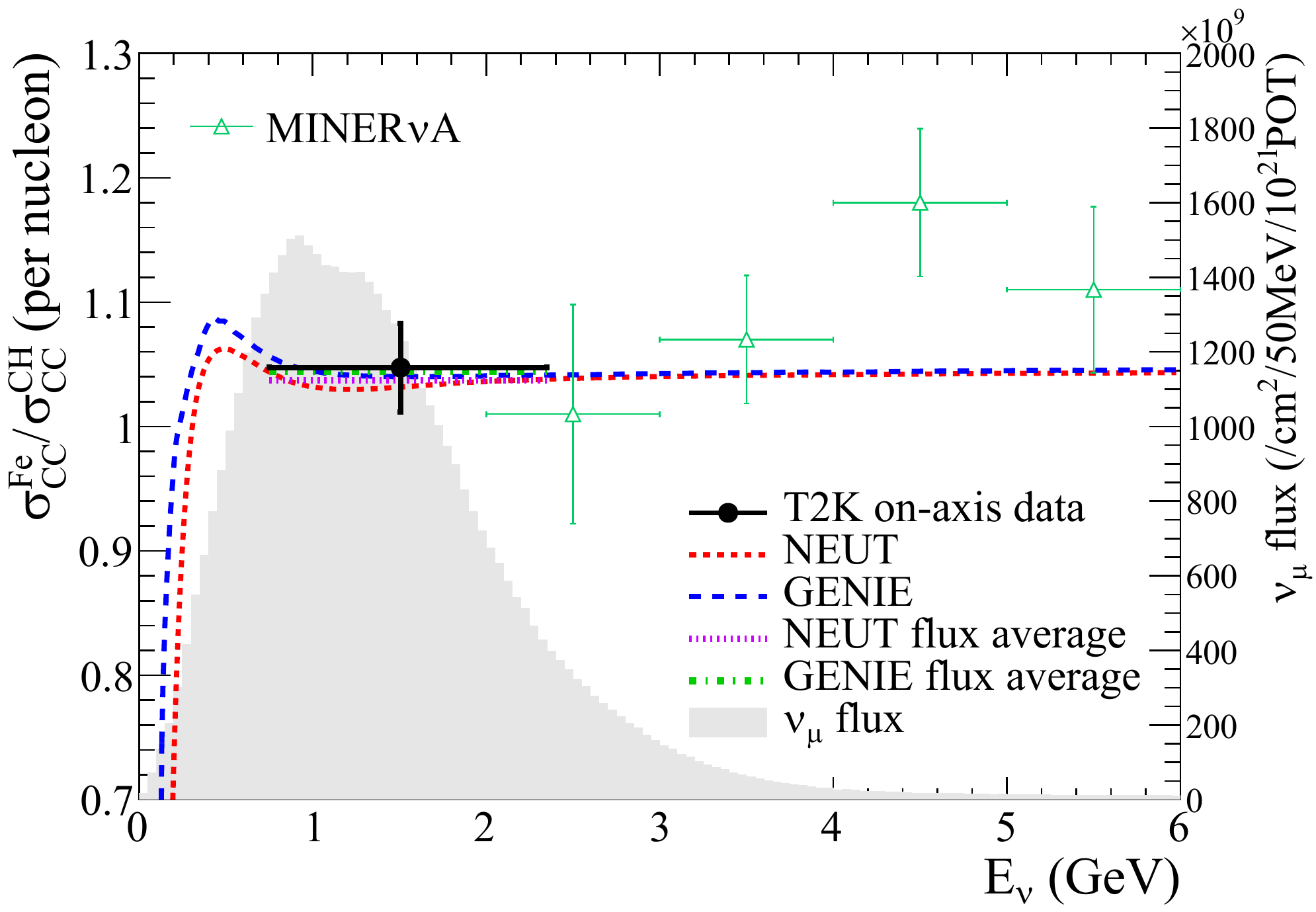}
  \caption{The inclusive $\nu_{\mu}$ charged current cross section ratio on Fe to CH with predictions by NEUT and GENIE. The isoscalar corrections are not applied to our data or predictions. Our data point is placed at the flux mean energy. The vertical error bar represents the total (statistical and systematic) uncertainty,
and the horizontal bar represents 68\% of the flux at each side of the mean energy. The MINER$\nu$A result is also plotted\cite{minerva_ratio}.}
  \label{cc_ratio_result}
  \end{center}
\end{figure}

The measured flux-averaged CC inclusive cross sections on Fe and CH and their ratio are
\begin{eqnarray}
\sigma_{\mathrm{CC}}^{\mathrm{Fe}}&=&(1.444\pm0.002(stat.)_{-0.157}^{+0.189}(syst.))\nonumber \\
&& \times 10^{-38}\mathrm{cm}^2/\mathrm{nucleon},\\
\sigma_{\mathrm{CC}}^{\mathrm{CH}}&=&(1.379\pm0.009(stat.)_{-0.147}^{+0.178}(syst.))\nonumber \\
&& \times 10^{-38}\mathrm{cm}^2/\mathrm{nucleon},\ \mathrm{and}\\
\frac{\sigma_{\mathrm{CC}}^{\mathrm{Fe}}}{\sigma_{\mathrm{CC}}^{\mathrm{CH}}}&=&1.047\pm0.007(stat.)\pm0.035(syst.),
\end{eqnarray}
at a mean neutrino energy of 1.51 GeV.
These are pure cross sections per nucleon for each atom, and isoscalar corrections are not applied.
They agree well with the predicted values from
NEUT and GENIE shown in Table~\ref{ccinc_prediction}.
The cross-section results are shown in Figs.~\ref{cc_incl_result} and \ref{cc_ratio_result} together with the predictions and measurements from other experiments.
Our result of the cross section ratio on Fe to CH is accurate to the level of 3\%. Hence, its consistency with the neutrino interaction models demonstrates that the target dependence of the nuclear effect is well understood and correctly treated in the models in 3\% level.

In Table~\ref{data_mc}, the measured CC inclusive cross section on CH and the ratios to the predictions by NEUT and GENIE are compared to those for the T2K off-axis neutrinos measured by the ND280 detector.
Here, it requires attention that the fluxes for these two detectors are highly correlated.
Both the ND280 and INGRID data are in good agreement with both the NEUT and GENIE models.

\begin{table}[htbp]
\begin{center}
\caption{CC-inclusive cross sections on Fe and CH and their ratio predicted by NEUT and GENIE.}
\begin{tabular}{cccc} \hline \hline
 & $\sigma_{\mathrm{CC}}^{\mathrm{Fe}}$ & $\sigma_{\mathrm{CC}}^{\mathrm{CH}}$ & $\sigma_{\mathrm{CC}}^{\mathrm{Fe}}/\sigma_{\mathrm{CC}}^{\mathrm{CH}}$\\ \hline
NEUT & 1.398$\times 10^{-38}$cm$^2$ & 1.348$\times 10^{-38}$cm$^2$ & 1.037\\
GENIE & 1.241$\times 10^{-38}$cm$^2$ & 1.188$\times 10^{-38}$cm$^2$ & 1.044\\ \hline \hline
\end{tabular}
\label{ccinc_prediction}
\end{center}
\end{table}

\begin{table}[htbp]
\begin{center}
\caption{The CC-inclusive cross section on CH measured with the T2K on-axis and off-axis fluxes and the ratios to the predictions by NEUT and GENIE. The errors represent the total (statistical and systematic) uncertainties.}
\begin{tabular}{ccc} \hline \hline
 & On-axis & Off-axis\\ \hline
Average energy & 1.51GeV & 0.85GeV\\
Data ($\times 10^{-38}\mathrm{cm}^2$)& $1.379_{-0.147}^{+0.178}$ & $0.691\pm0.085$\\
Data/NEUT & $1.023_{-0.109}^{+0.132}$ & $0.950\pm0.117$\\
Data/GENIE & $1.160_{-0.124}^{+0.150}$ & $1.057\pm0.130$\\\hline \hline
\end{tabular}
\label{data_mc}
  \end{center}
\end{table}

\section{Conclusions}\label{sec:conclusion}
We have reported the first neutrino cross section measurement with the T2K on-axis near neutrino detector, INGRID.
We have selected a sample of inclusive $\nu_\mu$ CC interactions in an INGRID standard module and the Proton Module.
From the number of selected events, the flux-averaged CC inclusive cross sections on Fe and CH and their ratio at a mean neutrino energy of 1.51 GeV have been measured.
These results agree well with the model predictions.

\section*{Acknowledgements}
It is a pleasure to thank Mr. Taino from Mechanical Support Co. for
helping with the construction of INGRID.
We thank the J-PARC staff for superb accelerator performance and the CERN NA61 collaboration for providing valuable particle production data. We acknowledge
the support of MEXT, Japan; NSERC, NRC and CFI,
Canada; CEA and CNRS/IN2P3, France; DFG, Germany; INFN, Italy; Ministry of Science and Higher Education, Poland; RAS, RFBR and MES, Russia; MICINN
and CPAN, Spain; SNSF and SER, Switzerland; STFC,
U.K.; and DOE, U.S.A. We also thank CERN for the
UA1/NOMAD magnet, DESY for the HERA-B magnet
mover system, and NII for SINET4. In addition participation of individual researchers and institutions has been
further supported by funds from: ERC (FP7), EU; JSPS,
Japan; Royal Society, UK; DOE Early Career program, U.S.A.

\bibliographystyle{apsrev4-1}
\bibliography{main}

\end{document}